\pdfoutput=1
\documentclass[traditabstract]{aa}
\usepackage{graphicx}
\usepackage{epstopdf}
\usepackage{subfigure}
\usepackage{color}
\usepackage{natbib}
\def\fdg{\hbox{$^\circ$}}

\begin{document}

\title{NELIOTA: The wide-field, high-cadence lunar monitoring system at
  the prime focus of the Kryoneri telescope}

\author{  
E.M.\, Xilouris\inst{1} \and
A.Z. Bonanos\inst{1} \and
I. Bellas-Velidis\inst{1} \and
P. Boumis\inst{1} \and
A. Dapergolas\inst{1} \and
A. Maroussis \inst{1} \and
A. Liakos\inst{1} \and
\\
I. Alikakos\inst{1} \and
V. Charmandaris \inst{1,2} \and
G. Dimou\inst{1} \and
A. Fytsilis\inst{1} \and
M. Kelley\inst{3} \and
D. Koschny\inst{4,5} \and 
V. Navarro\inst{6} \and
\\
K. Tsiganis\inst{7} \and
K. Tsinganos\inst{8}
}
\institute{	
Institute for Astronomy, Astrophysics, Space Applications \& Remote Sensing, 
National Observatory of Athens, P. Penteli, GR-15236 Athens, Greece
\and 
Department of Physics, University of Crete, GR-71003 Heraklion, Greece
\and
DFM Engineering, Inc., 1035 Delaware Avenue, Longmont, CO 80501, USA
\and
Scientific Support Office, Directorate of Science, European Space Research and Technology Centre (ESA/ESTEC),
2201 AZ Noordwijk, The Netherlands
\and 
Chair of Astronautics, Technical University of Munich, 85748 Garching, Germany
\and
European Space Astronomy Centre (ESA/ESAC), Camino bajo del
Castillo, s/n, Urbanizacion Villafranca del Castillo, Villanueva de la Ca$\tilde{\rm{n}}$ada, E-28692 Madrid, Spain
\and
Department of Physics, Aristotle University of Thessaloniki, GR-54124 Thessaloniki, Greece
\and
Section of Astrophysics, Astronomy and Mechanics, Department of Physics, University of Athens, 
Zografos, GR-15783 Athens, Greece
}	
   \offprints{E. M. Xilouris, \email{xilouris@noa.gr}}   
   \date{Received / Accepted }   
   \authorrunning{Xilouris et al.} 
   \titlerunning{The NELIOTA lunar monitoring system}
   % context, aims, methods, results, conclusions

\abstract{We present the technical specifications and first results of 
the ESA-funded, lunar monitoring project ``NELIOTA'' (NEO
Lunar Impacts and Optical TrAnsients) at the
National Observatory of Athens, which aims to determine the
size-frequency distribution of small Near-Earth Objects (NEOs) via
detection of impact flashes on the surface of the Moon. For the purposes of this 
project a twin camera instrument was specially designed
and installed at the 1.2 m Kryoneri telescope utilizing the fast-frame capabilities of
scientific Complementary Metal-Oxide Semiconductor detectors
(sCMOS). The system provides a wide field-of-view (17.0$^\prime \times$14.4$^\prime$) 
and simultaneous observations in two photometric bands (R and I), reaching limiting
magnitudes of 18.7 mag in 10 sec in both bands at a 2.5 signal-to-noise level. 
This makes it a unique instrument
that can be used for the detection of NEO impacts on the Moon, as well as for any
astronomy projects that demand high-cadence multicolor observations. 
The wide field-of-view ensures that a large portion of the Moon is observed, while the  
simultaneous, high-cadence, monitoring 
in two photometric bands makes possible, for the first time, the determination of the
temperatures of the impacts on the Moon's surface and the validation of the impact 
flashes from a single site.
Considering the varying background level on the Moon's surface we demonstrate 
that the NELIOTA system can detect NEO impact flashes at a 2.5 signal-to-noise level
of $\sim12.4$ mag in the I-band and R-band for observations made
at low lunar phases ($\sim 0.1$). 
We report 31 NEO impact flashes detected during the first year of the NELIOTA campaign.
The faintest flash was at 11.24 mag in the R-band
(about two magnitudes fainter than ever observed before) at lunar phase 0.32. 
Our observations suggest a detection rate of 1.96$\times10^{-7}$ events km$^{-2}$ h$^{-1}$.}

\keywords{Instrumentation: detectors - Techniques: miscellaneous -
  Telescopes - Moon - Surveys} \maketitle
%   \authorrunning{Xilouris et al.} 
%  \titlerunning{The NELIOTA twin fast-frame imaging system}   
%________________________________________________________________   

%%%%%%%%%%%%%%%%%%%%%%%%%%%%%%%%%%%%%%%%%%%%%%%%%%%%%%%%%%%%%%%%%%%%%%%%%
% Section 2
%%%%%%%%%%%%%%%%%%%%%%%%%%%%%%%%%%%%%%%%%%%%%%%%%%%%%%%%%%%%%%%%%%%%%%%%%
\section{Introduction}\label{INTRO}

Lunar monitoring provides a promising method for determining the size-frequency
distribution \citep[SFD;][]{Ivanov02, Werner02, Harris15} of small near-earth
objects (NEOs) via the detection of NEO impact flashes on the Moon
\citep[see][and references therein]{Bouley12, Suggs14}. Determining the SFD for
objects in the decimeter to meter range is important for evaluating the danger
of small NEOs impacting Earth, in light of the recent Chelyabinsk event
\citep{Brown13}, as well as the risk to artificial satellites and future
man-made stations on the Moon. The Lunar Reconnaissance Orbiter (LRO) provides an
indirect measurement of the SFD, by determining the cratering rate on the Moon
via temporal imaging \citep{Speyerer16}. So far, one new crater has been
associated with an observed impact flash \citep{Suggs14, Robinson15},
demonstrating the potential of the synergy between lunar monitoring from the
ground and space.

%%%%%%%%%%%%%%%%%%%%%%%%%%% FIG. 1
\begin{figure*}
  \centering
  \includegraphics[width=18cm,angle=0]{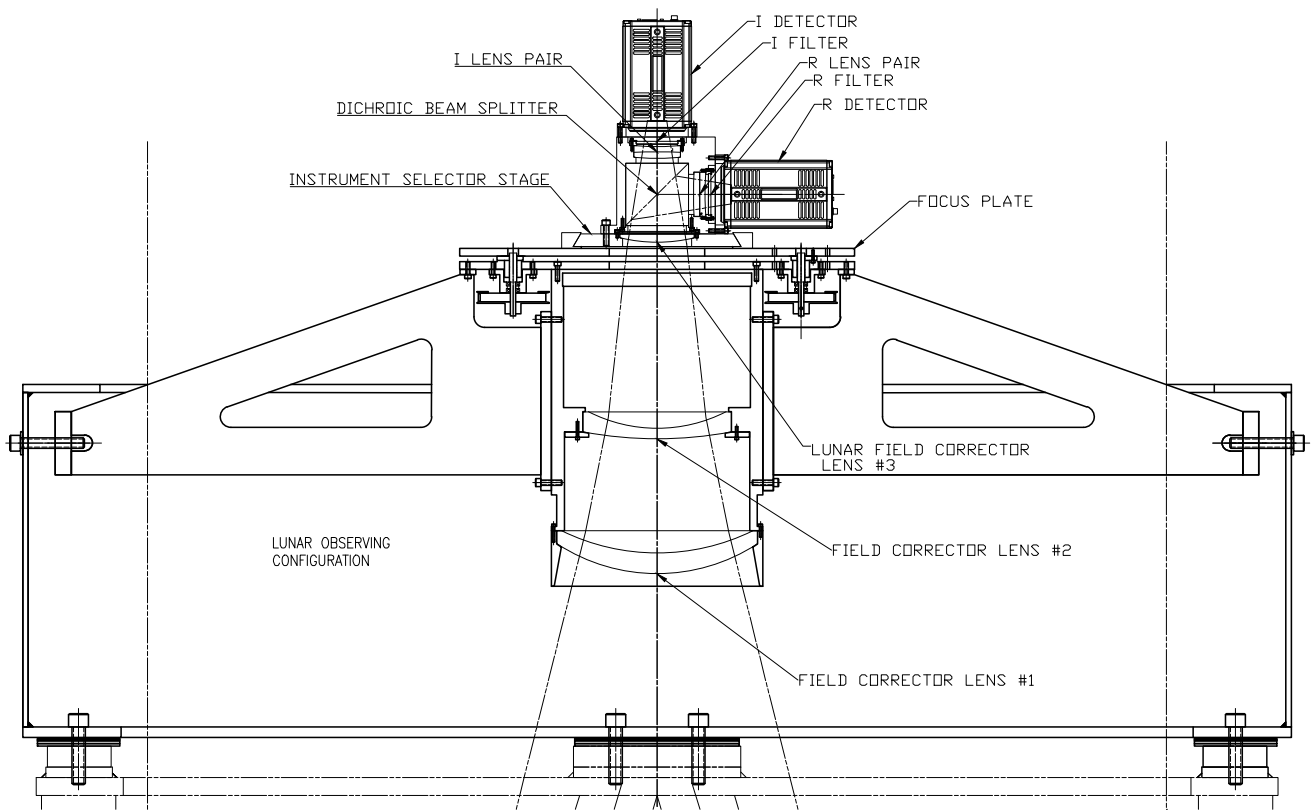}
  \caption{Section view of the Kryoneri Prime Focus Instrument (KPFI) optical layout. 
The position of the
    field corrector lenses \#1, \#2, and \#3, the dichroic beam
    splitter, as well as the lens pair elements in front of the Lunar
    imaging detectors and the R- and I-band filters are shown. The
    focus plate and the instrument selector stage are also indicated
    (Credit: DFM Engineering Inc.).}
\label{Fig:optlayout}
\end{figure*}

The estimation of the NEO size from lunar monitoring observations depends on the
assumed value of the luminous efficiency $\eta_\lambda$, which is defined as the
ratio of the measured luminous energy emitted at a particular wavelength
$\lambda$ to the total kinetic energy of the meteoroid. It therefore implies the
knowledge of the impactor velocity, which cannot be accurately known for
sporadic meteoroids. The luminous efficiency is best constrained for meteoroids
originating from specific, well-studied, meteor showers \citep{Ortiz00}.
\citet{BellotRubio00} have obtained values of $\eta \sim 2\times 10^{-3}$ from
lunar Leonids, while \citet{Moser11} report values between $1.2 \times 10^{-3}$
and $1.6\times 10^{-3}$ from lunar impacts involving Geminid, Lyrid and Taurid
meteoroids. Results of numerical simulations \citep{Nemtchinov98} predict $\eta$
to range from a lower limit of $10^{-6}-10^{-5}$ to an upper limit of $10^{-3}$,
whereas impact simulations modeling the high velocity of the Leonids yield
values of $\eta$ on the order of $10^{-3}$ to $2 \times 10^{-2}$
\citep{Artemeva01}. In any case, it is evident that lunar impact observations
during meteor showers are valuable for constraining the value of $\eta$.
\citet{Suggs14} report estimated sizes of their observed impactors - originating
from non-sporadic meteoroids - to be on the order of centimeters.

Ground-based lunar monitoring typically employs modest-sized aperture ($\sim30$
cm) telescopes, which survey on the order of $\sim$10$^6$ km$^2$ of the lunar
surface at a time. The advantage this method offers, as compared to monitoring
meteors in the Earth's atmosphere using all-sky cameras, is that the monitored
surface area is larger by two orders of magnitude. Several groups have therefore
undertaken lunar monitoring surveys for NEO flashes \citep[e.g.][]{Madiedo14,
Ortiz15, Rembold15, AitMoulay15}. Ground-based surveys routinely use two or more
telescopes often located at different sites in order to distinguish noise,
seeing variations, cosmic rays and satellite glints from real impact events. A
lunar impact event is confirmed when a simultaneous detection has been made
independently by two telescopes.

The ``NEO Lunar Impacts and Optical TrAnsients'' (NELIOTA)
Project\footnote{https://neliota.astro.noa.gr} is an activity launched by ESA at
the National Observatory of Athens (NOA) in 2015. It aims to contribute to the SFD of
small NEOs, by developing a system that will perform a 22-month lunar monitoring
campaign. NELIOTA brings three innovative aspects to the field of lunar
monitoring: (1) a single telescope with a twin camera system, which detects and
confirms events, (2) the twin cameras simultaneously monitor the Moon in two
photometric bands, allowing for a temperature determination of each flash and
the evolution of the flash, in case of flashes detected on multiple-frames
\citep{Bonanos17}, and (3) a large aperture telescope, which, in principle, can
detect fainter impact flashes than the modest-sized aperture telescopes
typically used. The limitation to the faintness of observed lunar impacts arises
from the variable brightness of the lunar background \citep[i.e.\
earthshine;][]{Goode01}, in combination with the wavelength observed and the
frame rate of the observations.

The selection of telescope and cameras were made so as to satisfy the main
project requirements: (1) to monitor the Moon at $\geq 20$ frames per second, 
(2) detect lunar
impacts down to 12$^{\rm th}$ magnitude, (3) have the ability to distinguish false
positives (e.g.\ cosmic rays, foreground glints) from real flashes, and (4) to
guarantee availability of the telescope for continuous usage that is better than
95\%.

%%%%%%%%%%%%%%%%%%%%%%%%%%% FIG. 2
\begin{figure}[t]
  \centering
  \includegraphics[width=9.cm,angle=0]{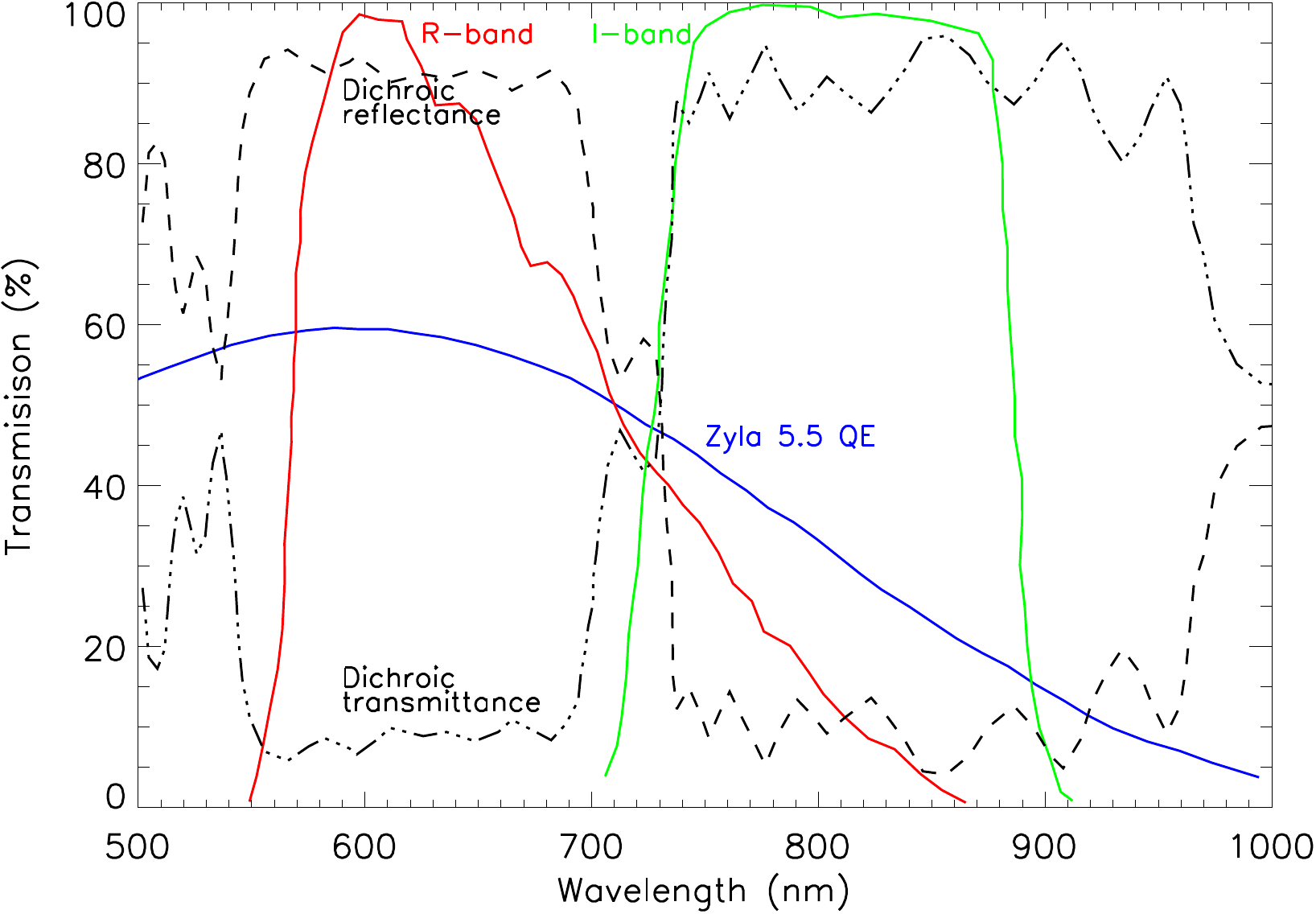}
  \caption{KPFI filters (red solid line for the R-band and green solid line
for the I-band) and dichroic beam splitter transmission curves (dashed and dashed-dotted
lines) combined with
    the QE response of the Lunar imaging detectors (blue solid line). Filter curves
    reflect the Johnson R- and I-band. The dichroic is centered at
    730 nm with a $> 90\%$ throughput.}

\label{Fig:filters}
\end{figure}

To address the specific needs of the NELIOTA project the 1.2~m Kryoneri 
telescope\footnote{http://kryoneri.astro.noa.gr}
of NOA was upgraded in 2016, commissioning a prime focus, high-speed, twin-camera 
Lunar imager.
The project has deployed a hardware system for recording and processing images.
We have also developed a software system, which controls both the telescope and
the cameras, processes the images and automatically detects candidate NEO lunar
impact flashes. The impact events are verified, characterized and made available
to the scientific community and the general public via the NELIOTA
website\footnote{https://neliota.astro.noa.gr/DataAccess} within 24 hours of
discovery. NELIOTA completed its commissioning phase in early 2017 and began a
22 month observing campaign in February 2017 in search of NEO impact flashes on the
Moon. The 1.2~m Kryoneri telescope is capable of detecting flashes much fainter
than all current, smaller-aperture, lunar monitoring telescopes. NELIOTA is therefore
expected to characterize the size-frequency distribution of NEOs weighing as
little as a few grams.

%%%%%%%%%%%%%%%%%%%%%%%%%%% FIG. 3
\begin{figure}[t]
  \centering
  \includegraphics[width=8.9cm,angle=0]{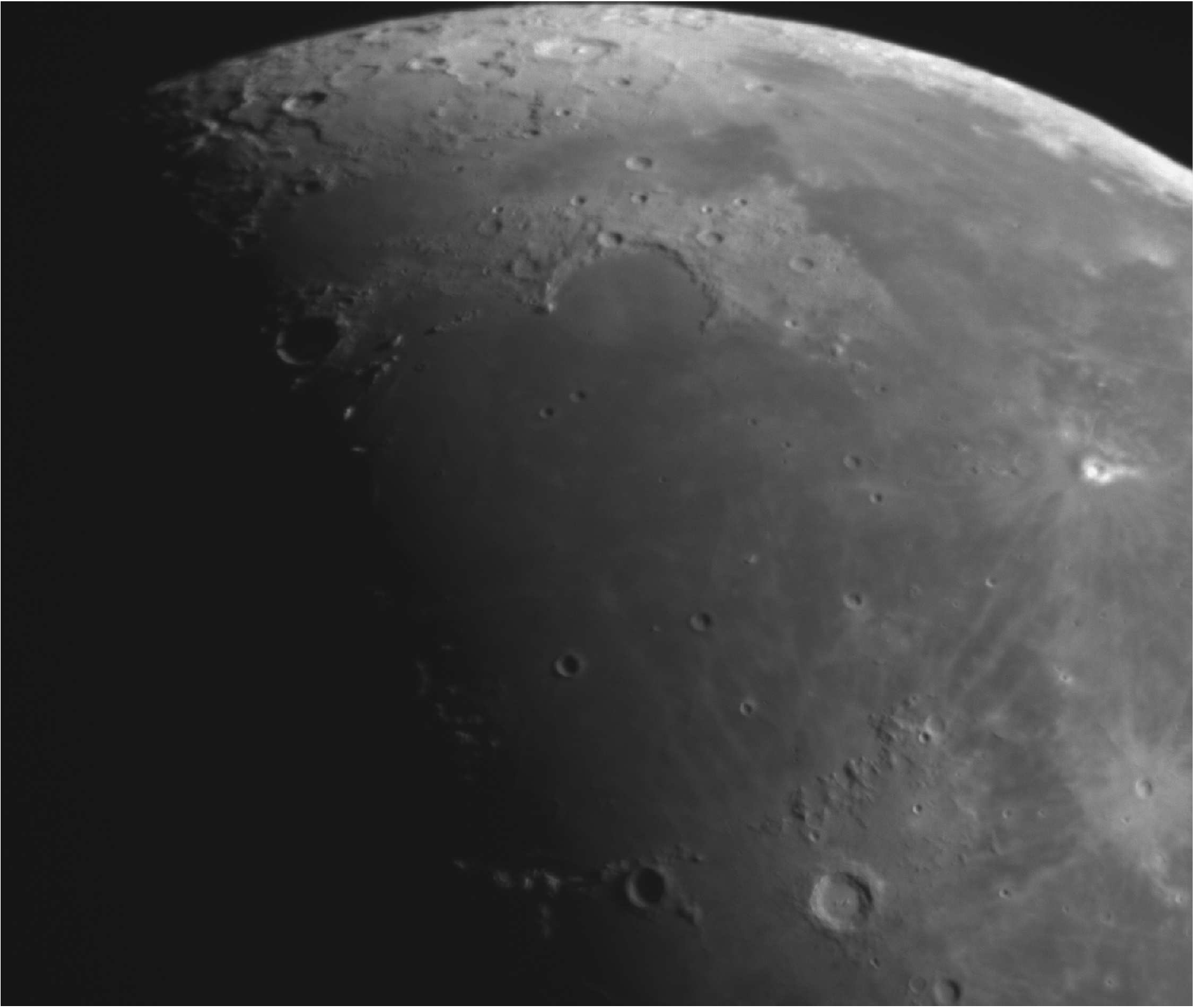}
  \caption{I-band image of the Moon observed with the NELIOTA
    system illustrating the available FOV, which is 17.0$^\prime \times$
    14.4$^\prime$.}
\label{Fig:Moon}
\end{figure}

\citet{Bonanos17} present the first scientific results of the project, in
particular, the first temperature measurements of impact flashes. This paper
presents a detailed description of the NELIOTA system and its components. Sect.~\ref{TELSYS} 
presents the upgraded 1.2~m Kryoneri telescope and its
new optical design, while the Lunar imager camera system is detailed in Sect.~\ref{CAMSYS}.
The NELIOTA system is presented in Sect.~4, while a brief description of the
system's software is given in Sect.~5.
The instrument's performance is evaluated in Sect.~6 with the results of the first year 
of the NELIOTA campaign presented in Sect.~7. Finally, a summary of the NELIOTA system 
and its first results are given in Sect.~8.

\section{The 1.2~m Kryoneri telescope}\label{TELSYS}

The 1.2 m Kryoneri telescope was selected (after a trade-off analysis) as the optimal 
facility of NOA for the NELIOTA project.
It was built by Grubb-Parsons and commissioned
in 1975. The telescope is situated at Kryoneri Observatory ($37^{\circ}
58\arcmin 19\arcsec$ North, $22^{\circ} 37\arcmin 07\arcsec$ East), in
the district of Corinth in the northern Peloponnese, Greece, at the top
of mount Kyllini, at an altitude of 930~m, close to Kryoneri village.

During its first $\sim40$ years of operation, the telescope's optical system consisted
of a paraboloidal primary mirror of 1.2 m diameter and f/3 focal ratio and a
hyperboloidal secondary mirror (31 cm). Both mirrors are made of Zerodur. This
configuration produced a final focal ratio of f/13 at Cassegrain focus. The main
scientific instrument in the last years has been a $2.5\arcmin \times
2.5\arcmin$ CCD camera Apogee Ap47p with a set of UBVRI filters. 

\subsection{The telescope retrofit}\label{TELSYSUPG}

The NELIOTA science objectives imposed strict requirements on the
optical design in order to provide: (i) the capacity to image
simultaneously in two bands, (ii) a large FOV ($\sim
20\arcmin$) and (iii) seeing-limited image quality (with $\sim
1.5\arcsec$ being a typical site seeing at Kryoneri Observatory
(typical dome seeing $\sim2.5\arcsec$), yielding a scale of $\sim 0.4\arcsec$ pixel$^{-1}$).

In 2016 the telescope underwent an extensive
upgrade by DFM Engineering Inc.\footnote{http://www.dfmengineering.com}, within 
the NELIOTA project. The
electro-mechanical upgrade included replacement of the telescope servo-motors
and associated hardware (including new encoder systems), and deployment of a new system 
for dome opening and rotation as well as motorized primary mirror doors. A computerized telescope 
control system (TCS), developed by DFM,
replaced the legacy console and allowed moving the observation control out of
the telescope housing area. A new, GPS-based time server (Meinberg, LanTime
M200/GPS) has been installed and is used by the TCS and other systems in the
Observatory. 

Furthermore, and according to the requirement for lunar observations, the
optics of the telescope were modified to operate with instruments at the prime
focus, bringing the telescope back to its primary mirror f/3 focal ratio and
providing an unvignetted field-of-view (FOV) of $\sim1.4$ degrees. A twin imaging
system, the Kryoneri Prime Focus Instrument (KPFI), designed and developed by
DFM Engineering Inc., is now in use, sampling 17.0$^\prime \times$ 14.4$^\prime$
of the total corrected FOV at the prime focus of the telescope, providing simultaneous 
high-cadence observations in two bands. 
With such a FOV, a significant fraction of the non-sunlit part of the Moon,
the only `usuable', for our purposes, can be
monitored to detect faint flashes coming from NEO impacts.
A direct imaging optical configuration, using a separate CCD detector, was also added to the
design to allow use of the full FOV. This was made possible through a
computer controlled camera slider plate mechanism allowing for two
operating modes, either using the twin imaging system (the ``Lunar
imager'' hereafter) or the direct imaging configuration (the ``Direct
imager'' hereafter).

%%%%%%%%%%%%%%%%%%%%%%%%%%%%%%%%%%%%55%%%%% TABLE 1
\begin{table}[t]
\tiny
\caption{Basic characteristics of the optical elements.}
\begin{center}
\begin{tabular}{lccc}
\hline\hline
Element & Material & Diameter& Thickness\\
        &          &(cm)&(cm)\\
\hline
Primary mirror & Zerodur& 120.0 &  - \\
Lens \#1 &N-BK7 &24.01 & 2.54\\
Lens \#2 &N-BK7 &17.78 & 1.27\\
Lens \#3 &N-BK7 &8.89 & 1.01\\
Lens pairs &N-BK7/SF4 &5.58 & 1.39\\
%Dichroic & & & \\
\hline
\end{tabular}
\end{center}
\label{Table:Lenses}
\end{table}

\subsection{The new Kryoneri Prime Focus Instrument (KPFI)}\label{TELSYSPFI}

Several optical elements were needed to overcome the challenge of
placing an instrument at the prime focus of the 1.2~m Kryoneri
telescope. The existing parabolic primary mirror (f/3) of the telescope
was used along with a set of field corrector lenses in order to focus
the beam in the prime focus. In total, three corrector lenses were used
(see layout in Fig.~\ref{Fig:optlayout} and characteristics in
Table~\ref{Table:Lenses}).
The first and second field corrector elements (labeled as field
corrector lens \#1 and \#2 in Fig.~\ref{Fig:optlayout}) are common for
the two imaging modes. This configuration produces a f/3.1 focal ratio
for the direct imager providing a maximum usable FOV of $\sim1.4$
degrees.  A third corrector lens (labeled as field corrector lens \#3 in
Fig.~\ref{Fig:optlayout}), followed by a dichroic beam splitter were
placed in the optical path of the Lunar imager producing a focal ratio
of f/2.8 in both channels. In front of each detector a lens pair is
placed in order to provide the necessary focal reduction and the same
effective focal length in each channel. A set of Johnson-Cousins R-
and I-band filters is used so that calibrated photometric measurements
can be made. The choise of this set of filters also assures direct
comparison with other NEO studies already conducted (e.g. \cite{Suggs14}).
Fig.~\ref{Fig:filters} presents the filter response
functions and the dichroic beam splitter transmission curves.

Field corrector lenses \#1 and \#2 were placed inside a cylindrical
baffle in order to reduce any scattered light. The cylinder is topped
with a structural plate that attaches to the spyder vanes.  Above the
structural plate is the focus stage (allowing for a focus travel of 13.3
mm) and above it, the camera selector stage. The camera selector stage
is motorized and remotely controlled, facilitating selection between the
two imaging modes.

The dichroic beam splitter was positioned 95.25 mm from the Lunar
imaging detectors (see Fig.~\ref{Fig:optlayout}). The specifications of
the dichroic were constrained by the beamsize and wavelength
distribution. The dichroic element consists of two 75 mm prisms. Prism
diagonals were coated to make a dichroic beam splitter that reflects 730
nm and shorter light (R-band) and allows light from 730 nm and
longer wavelengths to pass through (I-band). The prisms are aligned to
within $3\arcmin$. Appropriate anti-reflective coatings were applied to
the faces of the prisms. In order to avoid vignetting, we specified a
minimum clear aperture of $100.0 \times 142.0 \pm 0.5$~mm. The
reflected beam has a range of $530-730$ nm with a throughput of $>
90\%$, and the transmitted beam has a range of $730-950$ nm, with an
identical throughput to its counterpart (see Fig.~\ref{Fig:filters}).

%%%%%%%%%%%%%%%%%%%%%%%%%%%%%% TABLE 2
\begin{table}
\tiny
\caption{Specifications of the Andor Zyla 5.5 sCMOS camera for the
NELIOTA project with a 2$\times$2 binning configuration.}
\begin{center}
\begin{tabular}{ll}
\hline\hline
Parameter & Specification \\
\hline
Sensor type & Front illuminated Scientific CMOS \\
Pixel size & 6.48 $\mu$m \\
Active pixels & 1280$\times$1080 (binning 2$\times$2)\\
Size & 16.6$\times$14.0 mm$^{2}$ \\
Shutter & Global \\
Gain settings & Low \& High well capacity (16-bit) \\
Gain & 0.4 e$^{-}$ per A/D count\\
Read noise & 5.1 e$^{-}$ rms\\
Read-out rate & 560 MHz (280 MHz $\times$ 2 sensor halves) \\
Pixel scale & 0.8 arcsec pixel$^{-1}$  \\
& (or $\sim$1.5~km pixel$^{-1}$ on the lunar surface) \\
Frame rate & 30 fps \\
Field of view & 17.0$\times$14.4 arcmin$^{2}$ \\
Exposure time & 23 msec \\
Linearity & $<$ 60,000 ADUs (16-bit) \\
Cooling & Thermoelectrical (constant at 0$\fdg$ C) \\
Connection & USB 3.0 \\
\hline
\end{tabular}
\end{center}
\label{Tab:Zyla}
\end{table}

\section{The Lunar Imager camera system}\label{CAMSYS}

Systems designed, so far, for monitoring lunar impact events mostly use
black-and-white CCD video cameras. These are rather small size detectors (about 700
$\times$ 500 pixels) operating at 25 to 30 frames per second
rate producing interlaced 8-bit frames. In this way, the effective rate is doubled,
but a half frame (including either the odd or the even rows) is recorded after each
exposure. For the NELIOTA project large size sCMOS (scientific Complementary
Metal-Oxide Semiconductors) detectors are utilized. 

%%%%%%%%%%%%%%%%%%%%%%%%%%% FIG. 4
\begin{figure*}[ht]
 \centering
 \includegraphics[width=18.5cm]{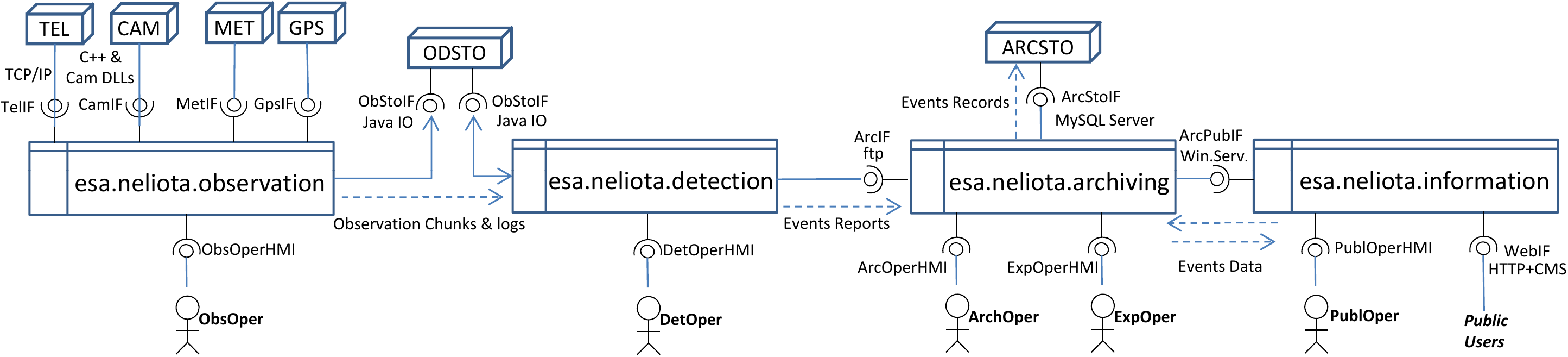}
 \caption{The NELIOTA system architecture and communication block diagram. The four software modules
(observation, detection, archiving and information), the basic hardware telescope (TEL), camera system (CAM),
time server (GPS), meteorological station (MET), high-performance storage (ODSTO), archiving storage (ARCSTO)
and their interfaces (IF), as well as the human-machine-interfaces (HMI) are shown.}
 \label{Fig:SysArch}
\end{figure*}

The twin imaging system for NELIOTA  includes a pair of identical
detectors (Zyla 5.5 sCMOS) providing simultaneous observations in two photometric 
bands. These detectors were selected since they satisfy
the project requirements for a fast-frame rate, high sensitivity and
resolution, and a light and compact design (being in the prime focus they had to be
small enough in size in order not to cause vignetting of light from the source)
making NELIOTA the first, to our knowledge, astronomical system to use
sCMOS detectors at a fast-frame rate. 
The active chip is 22 mm in diagonal with 2560 $\times$ 2160 square pixels (5.5
Megapixels), each 6.48 $\mu$m in size. With the NELIOTA setup (see 
Table 2) these cameras provide images at a rate of 30 frames per second,
offering as small as a sub-microsecond interframe gap. Their
quantum efficiency is 60$-$50\% in the R-band and 30$-$20\% in the I-band). 
Andor's Zyla 5.5 sCMOS offers a high-accuracy hardware-generated time-stamp on each frame,
essential for providing high precision time measurements of the detected lunar impact events.
For NELIOTA observations 
a 2$\times$2 binning configuration and a 30 frames per second rate has been chosen to optimize the
transfer rate of the obtained data. This configuration provides a pixel size
of 0.8 arcsec pixel$^{-1}$, well under the typical seeing measured on site, 
a FOV of 17.0$^\prime \times$14.4$^\prime$ (see Fig.~\ref{Fig:Moon}),
and also well sampled frames in the time domain. 

The sCMOS sensor has a highly parallel read-out architecture. Each of the 2560
columns possess a separate amplifier and analogue to digital converter, at both the
top and bottom of the column. While all columns are read out in parallel, the
read-out direction of each column is split in the center into the signal from the top
and bottom halves. There are two different pixel read-out rates, slow read at 200
MHz (100 MHz $\times$ 2 sensor halves) and fast read at 560 MHz (280 MHz $\times$ 2 sensor
halves), at either 12-bit or 16-bit mode, each with 3 different Gain settings
and two different read-out modes (Global and Rolling shutter). 
The mode selected for the NELIOTA
project is the fast read-out rate of 2$\times$280 MHz,
using the Gain settings of Low noise \& High well capacity (16-bit) and
the Global shutter. The latter offers the capability of all pixels being
exposed simultaneously and allows for easier synchronization of the
two cameras. The
specifications of the cameras are summarized in Table~\ref{Tab:Zyla}. An extensive evaluation 
of sCMOS cameras for astronomical purposes can be found in \cite{Qiu2013}. 

\section{The NELIOTA system} \label{SYSTEM}

The NELIOTA system has the ability to plan and ensure acquisition of the 
necessary data (Moon observations, calibration stars, flat-field and dark frames), to manage the data 
recording
to high capacity storage systems, to process the stored datasets and, finally, to detect possible 
impact events. Furthermore, the system organizes and
stores the detected events in the archive and, finally, provides useful 
information about the events on the dedicated NELIOTA web-based interface$^1$. 

The NELIOTA system is organized as a sequence of four functional
domains, namely, ``observation", ``detection", ``archiving", ``information", each of them controlling 
particular subsystems and assuring that the necessary information is shared among them
(see Fig.~\ref{Fig:SysArch}). 
The ``observation" domain controls the external hardware (telescope, cameras, GPS and 
meteorological station) and monitors their status, ensuring that the
desired observing plan is executed and that the raw (uncalibrated) frames and
metadata are stored in real time. The ``detection" domain provides 
post-observation data processing of the raw frames 
and applications of the impact detection algorithm (see Sect. 5.2). 
As a result of this process, candidate impact flashes are automatically detected and calibrated. 
They are then automatically forwarded to and treated within the ``archiving" domain, where 
they are stored in a database-driven archive,
and further evaluation and validation by expert scientists is performed. 
Finally, the ``information" domain ensures web-based publishing of all the relevant information
about the validated impact events and provides the general public with access to the NELIOTA 
results.
In addition to this, registered users have indirect, read-only, access (downloads) to the detected 
event files through the archiving system service, allowing for independent analysis of the datasets.
In particular, the registered user can download a data cube (in FITS format) of the detected
event accompanied by the seven frames of the observation before the beginning
of the event and after the end of the event so that a direct comparison of the local background
can be made.
For each observing run the detected events (if any) are published in the NELIOTA database$^3$ within
24 hours from the beginning of the observation.

\subsection{Design and implementation} \label{SYSIMPLEM}
Each domain includes a top-level software component which controls particular hardware subsystems through 
a dedicated machine-to-machine interface and/or processes the data flow as necessary. These operations
are controlled and monitored through a graphical user interface (GUI). Particular technologies have been 
used for the software development and implementation of the processing algorithms. With the exception of 
the ``information'' module, all others are developed in Java 8 SDK\footnote{https://www.oracle.com/java} 
platform with its widget toolkit ``JFC/Swing'' applied for creating the GUIs. The 
Eclipse\footnote{https://www.eclipse.org} Integrated Development 
Environment (IDE) has been used to write, to test and to package the three Java-based 
modules as JAR-executables.

%%%%%%%%%%%%%%%%%%%%%%%%%%% FIG. 5
\begin{figure}[t]
        \centering
        \includegraphics[width=9cm,angle=0]{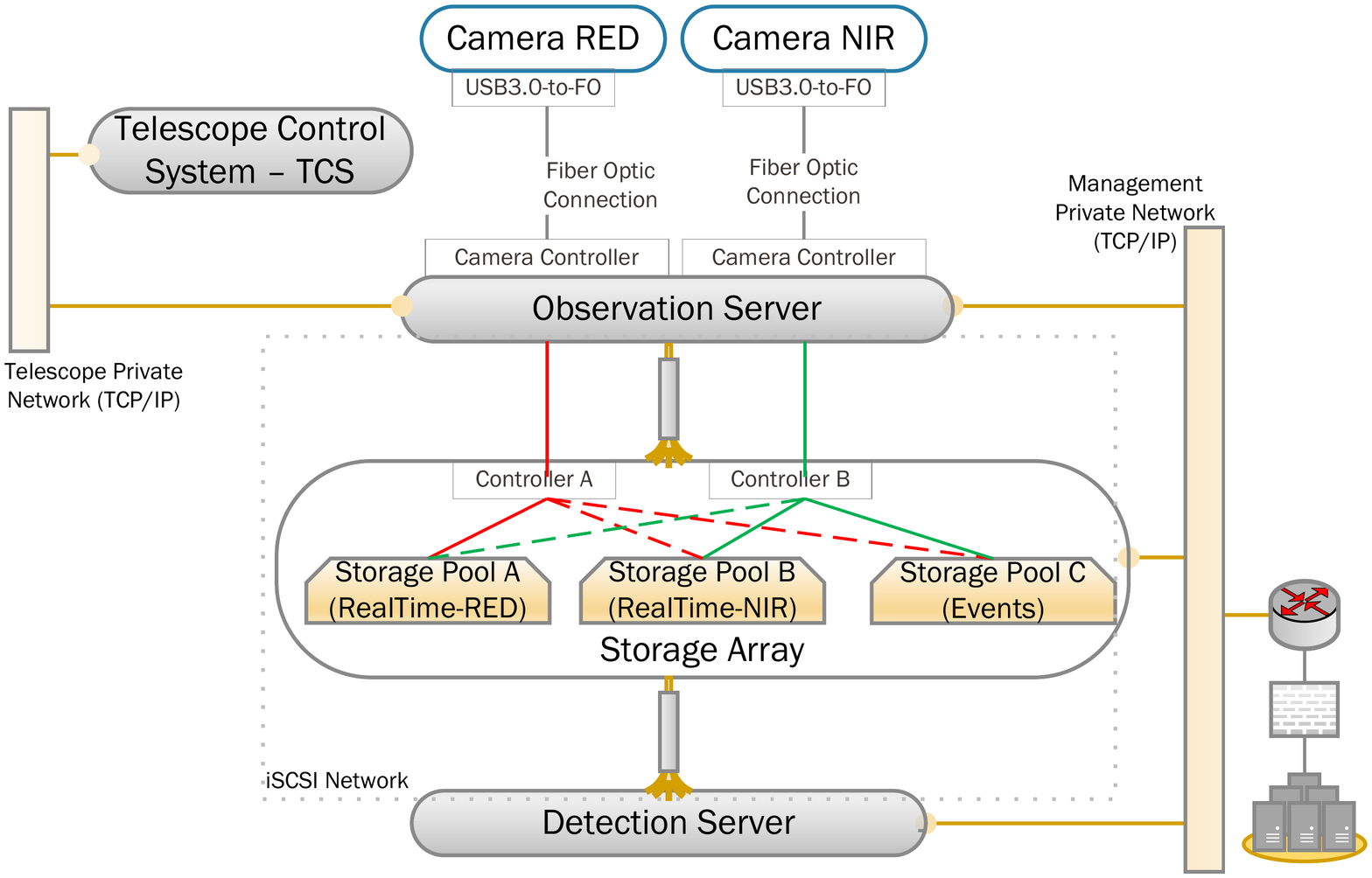}
        \caption{The configuration of the NELIOTA physical data system at the Kryoneri site.}
        \label{Fig:blockDiagram}
\end{figure}

The ``observation'' component is developed almost exclusively in Java, with a small interface 
in C++ to control, through a TCP socket, the two sCMOS cameras utilizing the supplier's library 
(Andor-SDK3). The telescope 
is controlled through a TCP/IP link to the Telescope Control System (DFM's WinTCS), exploiting the 
system's remote control commands. The GPS (ntp-service) and the weather
station services are passively accessed through corresponding NELIOTA system interfaces. The GUI 
utilizes two 4K ultra-HD monitors, one used for the operations control and for the monitoring
of the status of the NELIOTA system, and the other for real-time
display of the acquired frames. The observation system software is packaged in a executable JAR file.

The ``detection'' system is fully developed in Java and implements the algorithm for automated 
detection of the candidate impact events (see Sect 5.2), their characterization and
location on the Moon as well as the re-formatting of the raw frames into FITS format. 
An interface based on a FTP-client then transfers the detected event files to 
the remote archiving system. 

The ``archiving'' system (also written in Java) includes an interface to the FTP-server that allows
the data of the candidate impact events to
be transfered to the data-base driven archive (Microsoft SQL Server). Its GUI provides the appropriate 
tools so that the expert scientist can visually inspect the data and validate the detected 
events. Furthermore, the ``archiving'' system utilizes a Windows Service and a Web Service for interfacing
with the ``information'' domain. Through this interface the bulk of information of the new events are 
being transfered in a proper format, ready to be published on the web. 
The archive also, indirectly, services users requests to the ``information'' system for events, 
supplying the latter with properly packaged event records for further downloading. The executable JAR is 
activated as a non-stop service.  

The ``information'' system's top-level component is a website which implements its 
functionality as an ASP.net\footnote{https://www.asp.net} web application based on the Model-View-Controller 
(MVC) architectural pattern. Most components are written in C\#. The application registers itself on a
Microsoft IIS server to listen for public user requests to provide html-pages and to service downloads. 
It is the only domain to which public users have limited access. The system implements a Windows Service 
to manage downloading requests and a Web Service to get the necessary content from the 
archiving system. 

\subsection{NELIOTA system deployment} \label{SYSDEPLOY}
The functional requirements set above for the NELIOTA system (Section~\ref{SYSTEM}) impose 
special characteristics to the hardware in order 
(1) to assure synchronized operation of the telescope and the two cameras, 
(2) to achieve high performance storage capabilities in terms of speed, capacity, security and 
hardware-failure safe operations, 
(3) for storage virtualization, and 
(4) for high-performance data processing resources. 
For practical reasons that have to do with the location of Kryoneri Observatory, the NELIOTA
system had to be split into two subsystems in order to be most efficient. 
One of the subsystems is physically present at Kryoneri 
Observatory and the other is hosted at the computer center of IAASARS/NOA located in the northern 
suburbs of Athens, with the two of them connected through a private 11~Gbps RF-link. 
The ``observation'' and the ``detection'' systems are deployed at the Kryoneri site on a cluster of 
two nodes, both using a common external high-performance storage array of a total of 38.4~TB through a dedicated 
switch (Table~\ref{Tab:DeployKryoneri}). The storage ``subsystem'' is linked through 8$\times$1~Gb copper 
connections to the switch, while the two servers are connected to this switch by their optical SFP ports. These 
connections form a dedicated ring, i.e., an isolated high-speed iSCSI network 
for data transfer between the servers 
and the storage. 

%%%%%%%%%%%%%%%%%%%%%%%%%%%%%%%%% TABLE 3
\begin{table}[t]
\tiny
\caption{Specifications of the NELIOTA hardware systems at the Kryoneri site.}
\begin{center}
\begin{tabular}{ll}
\hline\hline
\multicolumn{2}{c}{Specifications of the two servers (per server)} \\
\hline
Model 	& HP ProLiant DL380 Generation9 \\
Processor 	& Intel Xeon E5-2660v3 CPU \\
 		& (2.66GHz, 10 cores, 20 threads, 25MB Cache) \\
Memory	& 64GB RAM (8$\times$8GB DDR4 PC4-2133P-R) \\
Storage 	& 2$\times$300GB (HP300 SAS 10K SFF) disks \\
Controller	& Smart Array P440ar/2G FIO RAID \\
Interface	& HP Ethernet 10GbE 2-port 530SFP+\\
System	& Windows Server 2012 R2, x64 \\
\hline
\multicolumn{2}{c}{High-performance disk-array specifications} \\
\hline
Model 	&  HP MSA 2040 SAN Storage\\
Storage	& 32$\times$1.2TB (HP1200 6G SAS 10K SFF hot-plug)\\
Controller	& 1GbE/10GbE iSCSI SAN \\
\hline
\multicolumn{2}{c}{Interconnecting switch specifications} \\
\hline
Model	& HP 2920-24G switch\\
Modules	& 2$\times$10GbE (HP 2020 2-port SFP+) \\
\hline
\end{tabular}
\end{center}
\label{Tab:DeployKryoneri}
\end{table}

A second ring is implemented as a typical communication network for device management, routing 
configuration and remote access to devices for healthy operation monitoring. 
The main service provided within this ring is the control of the telescope as well as the dome
of the building housing the telescope via the dedicated telescope 
control system (TCS, see Section~\ref{TELSYSUPG}) and through a 
dedicated TCP/IP network. The two cameras are connected to the observation server via optical links 
exploiting in full the USB 3.0 capability of each camera (see Section~\ref{CAMSYS}). The server 
passively accesses the remote meteorological system gathering environmental information. Both servers, 
like the TCS, are time-synchronized through the ntp-protocol by the local, GPS-based, 
server (Section~\ref{TELSYSUPG}). Both the ``observations'' and ``detection'' systems, are mirrored 
on the two nodes, so in the case of a failure, only the reconnection of the camera USB links from one 
node to the other is needed for resuming the observation. The NELIOTA physical data system at Kryoneri 
is presented in Fig.~\ref{Fig:blockDiagram}.

%%%%%%%%%%%%%%%%%%%%%%%%%%% FIG. 6
\begin{figure}[t]
  \centering
  \includegraphics[width=9cm,angle=0]{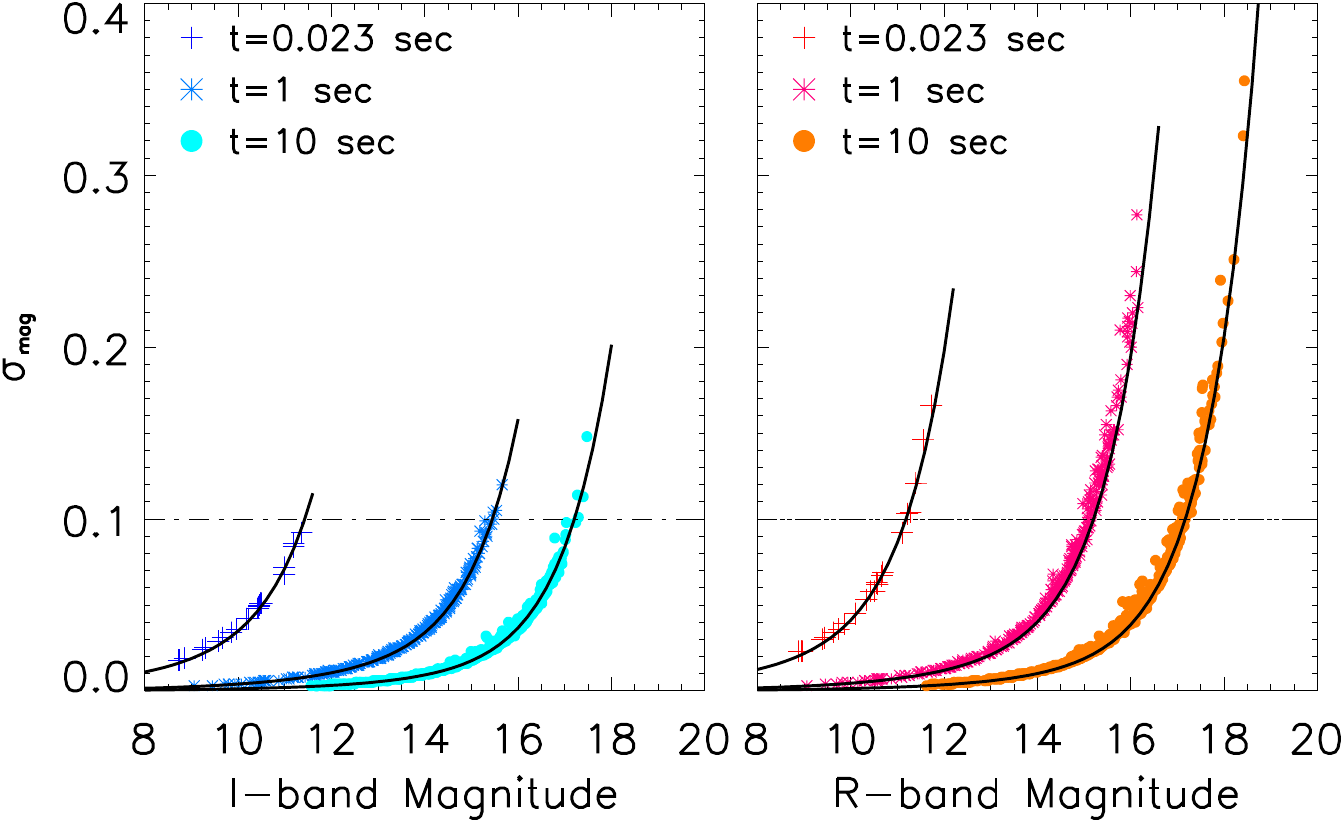}
  \caption{Instrument performance for the two channels (I- and R-band;
left and right panels respectively). In each panel the measured magnitudes of
stars in the open cluster NGC~1960 are plotted
against their measured uncertainty ($\sigma_{mag}$) for three different
exposure settings [10 sec (circles), 1 sec (asteriscs) and 0.023 sec (crosses)].
The curves passing through the photometric measurements are the model predictions
of Eq. 1 providing information on the magnitude level that can be reached at a certain $\sigma_{mag}$ level
(see text for more details). The horizontal line in each panel indicate the 0.1 mag noise level.}
\label{Fig:sigmamag}
\end{figure}

%%%%%%%%%%%%%%%%%%%%%%%%%% TABLE 4
\begin{table}
\tiny
\caption{Specifications of the NELIOTA virtual server hardware systems.} 
\begin{center}
\begin{tabular}{ll}
\hline\hline
\multicolumn{2}{l}{Virtual server specifications} \\
\hline
Processor       & QEMU V-CPU v. 1.1.2 2.27GHz \\
Memory  & 5GB (Archiving server), 8GB (Information server) \\
Storage         & 100GB (Archiving server), 50GB (Information server)  \\
System  & Windows Server 2012 R2, x64\\
\hline
\end{tabular}
\end{center}
\label{Tab:DeployAthens}
\end{table}

%%%%%%%%%%%%%%%%%%%%%%%%%%% FIG. 7
\begin{figure}[t]
  \centering
  \includegraphics[width=9cm,angle=0]{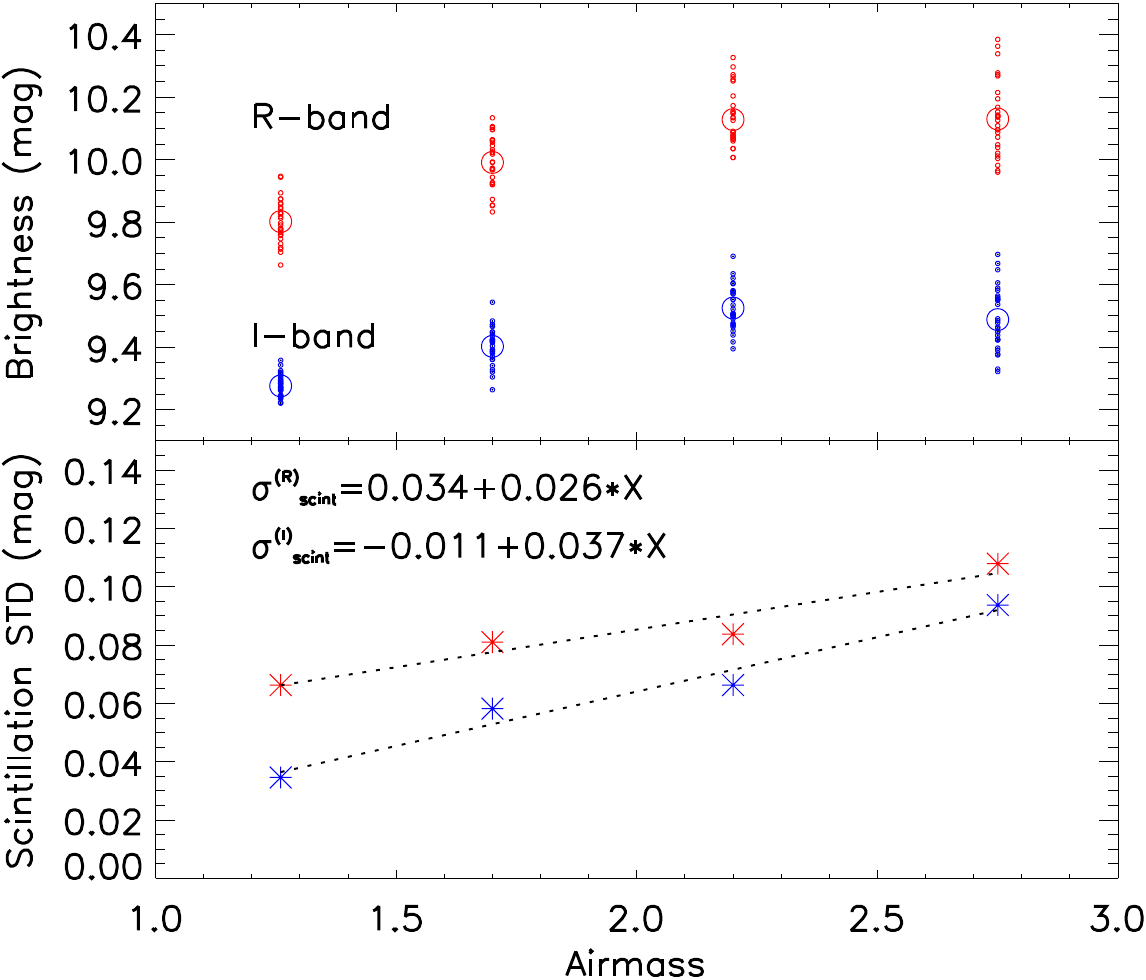}
  \caption{Scintillation noise ($\sigma_{scint}$) as a function of airmass.
The top panel shows the scatter of the instrumental R- and I-band brightness of
a photometric starndard star (SA~113-475) observed with an exposure time of 23 msec
at four different airmasses ($X$=1.26, 1.7, 2.2, 2.75;  small circles) along with the
mean values at each airmass (large circles). In the bottom panel the standard deviation
of the brightness of the source, in each airmass, is plotted against the
airmass with the linear fits to the data provided (see the text for more details).}
\label{scintillation}
\end{figure}

The NELIOTA ``archiving'' and ``information'' systems are deployed in the IAASARS/NOA
node on two virtual servers. The two servers are almost identical apart from the size of the dedicated 
memory and storage (Table~\ref{Tab:DeployAthens}). The hardware of the server system has been 
implemented on the Virtual Machines Cluster of the Network Information Center (NIC) of the National 
Observatory of Athens. 

\subsection{The data acquisition} \label{SYSPACQUIS}
The built system configuration achieves a throughput of 5 Gbps/camera (camera to storage array). 
The two controllers of the storage array have been programmed to operate in parallel and 
independently manipulate the two streams/cameras, routing the related data streams to the storage array 
(red and green solid lines in Fig.~\ref{Fig:blockDiagram}). Following this concept the storage array is 
divided into two parts (storage pools ``A'' and ``B'' organized in RAID-5 parity) and each controller 
directs the related stream to a separate set of physical hard disks. In addition, each controller operates 
as a backup of the other one ensuring the maintenance of operation in cases of hardware failure 
(dashed lines in Fig.~\ref{Fig:blockDiagram}). On the other hand, a storage pool ``C'' has been created
with RAID-1 parity in order to secure the data with the candidate events, produced by the detection system. 

%%%%%%%%%%%%%%%%%%%%%%%%%%% FIG. 8
\begin{figure*}[t]
\centering
\includegraphics[width=18.5cm,angle=0]{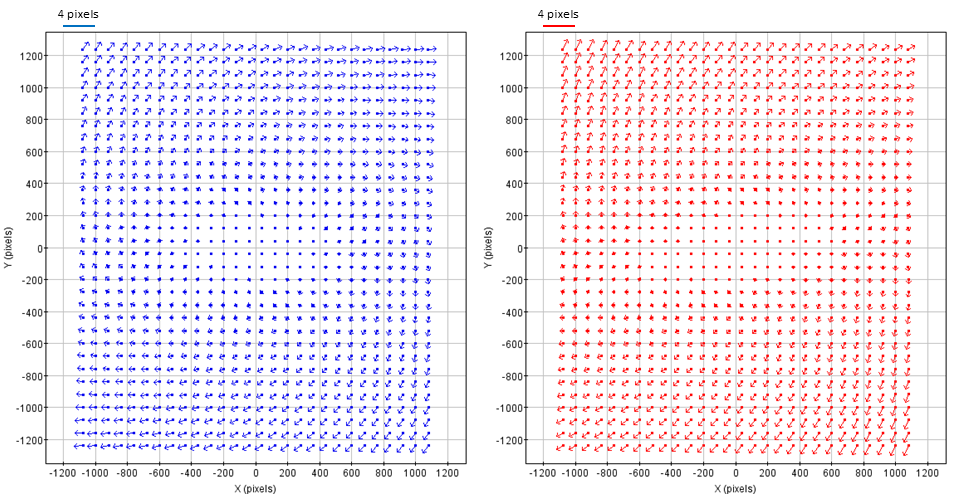}
  \caption{Field distortion map of the I-band (left panel)
and the R-band (right panel) frames. Each arrow shows the expected position
on the sky (beginning of arrow) and the measured position (end of arrow).
For displaying purposes the arrows are artificially enlarged by a factor
of 50 with a scale indicator (4 `enlarged' pixels) shown in the upper left
part of each plot.}
\label{Fig:distortion}
\end{figure*}

Based on the observation setup with the two cameras each acquiring 30 frames per second with 
a 2$\times$2 binning, the produced data rate is almost 1.4~Gbps and hence, the required storage capacity 
is about 612 GB/hour. It has been estimated that no more than 2.5-3.0 TB storage for raw data is 
required per one observing cycle of NELIOTA. Having in mind the requirements for observing
conditions (see Section~\ref{SWOBSERV}), the capacity of the storage allows to keep observed raw frames for 
at least two months. On the other hand, the storage pool ``C'' capacity provides enough storage to hold 
detected event records for a few years of NELIOTA operation. 

\section{The NELIOTA software}

A dedicated software system has been developed in order to
control the telescope, schedule observations, reduce individual frames,
and perform automatic detection of the impact events. In addition, an
entry in our web-based, user-friendly, database is automatically created
when a new event is detected (within 24 hours of the event
observation).  The software (presented in detail elsewhere; Fytsilis et
al. in prep) has been developed in JAVA using ECLIPSE IDE and
JAVA RE8 and is running on a Windows Server. It provides a computer
interface between all individual components of the NELIOTA system.

\subsection{The telescope/cameras control software} \label{SWOBSERV}

The NELIOTA observations are scheduled and performed using the
observation planner and acquisition functions of the software. The observation
planner produces an observing plan for each night according to the
visibility of the Moon and the observability of the photometric
calibrators. The observations are performed in the non-sunlit part of
the Moon when the illuminated fraction is between $\sim 10-45\%$
(i.e.\ from waxing crescent to the first quarter and from the last
quarter to the waning crescent phase). An elevation limit of
$20^{\circ}$ above horizon is set due to the dome clearance. The Moon is
observed at a frame rate of 30 frames per second for a 15 minute
interval or ``chunk'' before this dataset is stored on the server. A
selection of photometric calibrator stars from \citet{landolt} is made
by the planner according to their proximity to the Moon and observed in
between lunar observing ``chunks''. The exposure time of the calibrators
ranges from 0.05 to 4 sec depending on their apparent magnitude. 
In the case of an impact event the
calibrator observed closest in time to the event is selected to conduct
the photometric analysis. Ancillary observations (flat-field and
dark frames) are obtained on each night. Sky flat-field
frames are observed during twilight with exposure times set
automatically by the software (typically ranging between 0.04 to 2
sec) in order to make the best use of the dynamical range of the
detectors. Dark frames are obtained before and after
the end of lunar observations. The NELIOTA software commands the
telescope through the TCS interface, while the cameras are controlled by
using the software development kit (SDK) version 3.11 provided by Andor.

%%%%%%%%%%%%%%%%%%%%%%%%%%% FIG. 9
\begin{figure*}[t]
        \centering
        \includegraphics[width=18cm,angle=0]{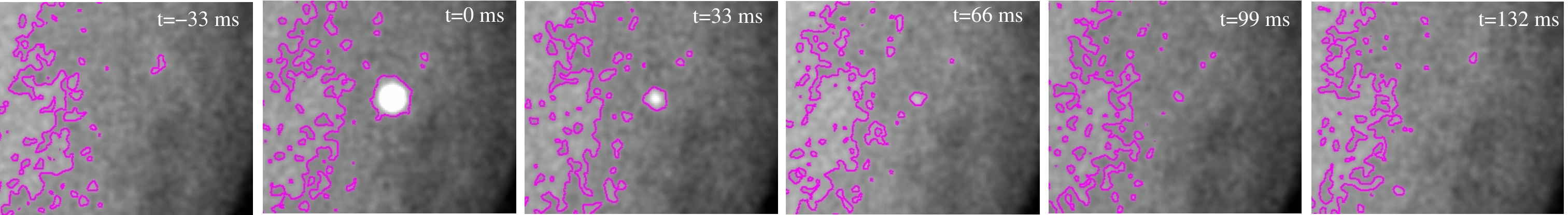}
        \caption{Time sequence of the I-band brightness of the impact detected
by the NELIOTA system at 04:35:09.967 UT on December
14$^{th}$, 2017 (event ID 30). The magenta contours show the 500 ADUs level, corresponding
to $\sim$2 signal-to-noise detection threshold. The leftmost and rightmost panels
are the frames before and after the event and are also displayed for comparison
purposes. The time of the beginning of the exposure of each frame is also
indicated in each panel (with the start of the event occurring at t=0 msec). The size
of each panel is 110 $\times$ 85 arcseconds.}
        \label{Fig:mosaic}
\end{figure*}

\subsection{The impact detection software} \label{detection}

Due to the large amount of data collected, a pipeline has been developed
in order to detect candidate impact events, which are later evaluated by
an expert scientist. The pipeline is based on two successive steps: (i)
calibration of the raw data, and (ii) application of the detection
algorithm.

At the observed rate of 30 frames per
second the actual on-source integration time is 0.023 sec followed
by a read-out time of 0.010 sec. Events that are detected in a
single frame are considered to have a duration of 0.033 sec (upper
limit), while for the events that are detected in more than one frames,
the total duration is calculated by summing the integration and read-out
times of the successive frames until the end of the event.

Calibration of the observed frames is done in a standard way by
performing a dark current subtraction using the median-combined dark
frames and also flat-field corrections using the median-combined sky
flat-frames observed. The calibrated lunar frames are then background
subtracted to remove the inhomogeneous surface of the Moon and
earthshine. To calculate the background $B_t$ for each individual frame
$I_t$ observed at time $t$, an adaptive background is created by
combining all individual frames of the Moon observed until $t-1$ (the frame 
observed before the one at time $t$). These
frames are weighted in time as $B_t=\alpha \times I_{t-1} + (1-\alpha)
\times B_{t-1}$ with $\alpha$ having values in the range (0, 1) depending
on the degree of importance of the most recent frames. The value that we
use in the NELIOTA setup is $\alpha=0.35$. This ensures that at the
observed rate of 30 frames-per-second, the main contribution in the
background comes from the most recent observations (closer to time $t$),
which reduces the effects of seeing as well as possible telescope
tracking deviations, but on the other hand produces a robust estimate of
the global background.

The background subtracted image of interest $D_t=I_t-B_t$ is then
subjected to a threshold detection process using a high and a low
threshold value. The high threshold value is primarily used to define
pixels with large deviations from the local background, while a low value
is subsequently used to better define these deviations by looking at the
fainter levels. When a candidate event is flagged, the connected
components are extracted so that the event can be treated as a single
object and its extent can be defined. With this criterion, objects that
are smaller than a minimum size (10 pixels) are considered to be noise
and are thus rejected. Objects that are large enough are then
considered as candidate events and are recorded. This same procedure is
then repeated for the subsequent frame ($t+1$) with the exception that
in the calculation of the new background, the area extracted by the
connected component analysis is not considered. This secures detection
of the same event in subsequent frames as the event fades.

For each candidate event that is detected by the software, visual
inspection is performed in order to validate it. The twin camera
system can easily rule out cosmic rays since they only appear in one of
the two frames. Other artifacts, such as satellites or airplanes
crossing the field-of-view can also be easily ruled out due to their
elongated trajectory. Once an event is validated, photometric
calibration is performed by an expert scientist. The photometric calibrator
closest (in time and angular distance) to the observation of the event is then
used to derive the photometric zero-point and thus calculate the
apparent magnitude of the event.
Events that are only detected in the I-band resembling flashes are
also stored and flagged as ``suspected''. 

The location of the flashes on the Moon is derived in a semi-automatic way
using a dedicated tool of the NELIOTA detection software and a high
detailed image of the Moon from the  Virtual Moon Atlas 6.0
(VMA6)\footnote{http://ap-i.net/avl/en/start} software. Using
the options available for the latter (i.e. date, time observing site, libration) it is feasible
to export a highly detailed image of the Moon as seen from the observing site at the exact
time when the flash is detected. The localization tool
stacks the images contained in the data cube of each event (see Section~\ref{SYSTEM}) in order
to produce a higher contrast image of the Moon in which the lunar features can easily
be spotted. Subsequently, the user imports the detailed image produced by the
VMA6 and identifies at least three lunar features (e.g. small craters) by
selecting them in both the VMA6 image and the stacked image. After the successful cross match,
the software exports the initial
VMA6 image in which the location of the flash is marked. The exact selenographic coordinates can
then be extracted.
The error in the location determination, taking into account the typical seeing
values for the site and also the error in corss matching features on the Moon between the
observations and the VMA6 map, is roughly estimated to be $6.5\arcsec$
($\sim 0.5$ degrees in selenographic coordinates).

All the information described above is made available on the
NELIOTA website$^1$ within 24 hours after the observation.

\section{Instrument performance}
 
\subsection{Noise characterization} \label{noise}
We have performed observations of the open cluster NGC~1960 
(RA$_{2000}$: $05^h 36^m 17^s$, DEC$_{2000}$: $+34^{\circ} 08\arcmin 27\arcsec$) in order to assess 
the instrument performance in terms of depth and instrument/photonic noise. 
The open cluster was observed simultaneously in the two bands 
on the night of March 11$^{th}$, 2018
with exposure 
times of 10, 1, and 0.023 sec (the last being the fixed exposure time setting for the NELIOTA 
observations) and four different airmasses ($X=1.027, 1.138, 1.598, 2.397$). 
For the photometry we used the {\it phot} task in IRAF\footnote{IRAF is distributed 
by the National Optical Astronomy Observatories,
which are operated by the Association of Universities for Research
in Astronomy, Inc., under cooperative agreement with the National
Science Foundation.}.
The photometry was performed within an aperture of 3 pixels radius, corresponding to 
$2.4\arcsec$,
while the sky background was measured in an annulus centered at each star with
an inner radius of 10 pixels and an outer radius of 15 pixels.
In each 10 sec exposure we used 20 relatively bright and isolated stars in the
field to calculate the aperture corrections needed for an aperture of 10 pixels
in order to account for the stellar light
missing in the smaller, 3 pixel, aperture used in the photometry of the cluster.
The mean sky background standard deviation ($STD_{bkg}$) was measured to be
12, 14, 38 counts in 0.023, 1, 10 sec, respectively, in the I-band and 12, 13, 30 counts 
in 0.023, 1, 10 sec, respectively, in the R-band. 
The uncertainty on the brightness of the source ($\sigma_{mag}$) was then 
calculated as 
\begin{equation}
\sigma_{mag}=1.0857/SNR
\end{equation}
with the signal-to-noise ratio ($SNR$) defined as:
\begin{equation}
SNR=\frac{N}{\sqrt{\sigma_{source}^2+\sigma_{bkg}^2+\sigma_{RN}^2}}
\end{equation}
(see, e.g., \cite{everett}).
Here, $N$ is the source flux in electrons, $\sigma_{source}$ the
source shot noise, $\sigma_{bkg}$ the background noise and $\sigma_{RN}$
the read-out noise. 
In particular, 
\begin{equation}
N=t\times G \times 10^{-0.4(mag-zpt)}
\end{equation}
with $t$ being the exposure time (in sec), $mag$, the magnitude of the source and $zpt$ the zero point in the
magnitude scale (see below for the computed values in the R- and I-bands ($r_0$ and $i_0$ respectively)).
The source shot noise is defined as $\sigma_{source}=\sqrt{N}$, the background noise as
$\sigma_{bkg} = STD_{bkg} \times G \times \sqrt{n_{source} (1+n_{source}/n_{bkg})}$ and the read-out noise as $\sigma_{RN} =
RN \times \sqrt{n_{source} (1+n_{source}/n_{bkg})}$. 
In this case $G=0.4$ e$^{-}$ per A/D count, and $RN=5.1$ e$^{-}$ 
(see Table~\ref{Tab:Zyla} for the specifications of the detectors),
while $n_{source}$ is the number of pixels inside the photometry aperture (radius of 3 pixels) and 
$n_{bkg}$ the number of pixels inside the background annulus (inner radius of 10 pixels and
outter radius of 15 pixels). Given the values above, $n_{source}/n_{bkg}=0.072$.
For each exposure setting, several isolated and well detected stars were chosen as photometric calibrators
for the rest of the stars in the cluster. In particular, 5, 20 and 26 such
stars were selected for the 0.023, 1 and 10 sec exposures
with their reference magnitudes 
obtained from \cite{jeff}. The above procedure resulted in 
the determination of the zero points in the magnitude scale ($r_0$, $i_0$), of atmospheric extinction 
coefficients ($r_1$, $i_1$) and of the color terms ($r_2$, $i_2$) in the following photometric equations: 

\begin{displaymath}
r = R + r_0 + r_1 X + r_2(R-I)
\end{displaymath}
\begin{displaymath}
i = I + i_0 + i_1 X + i_2(R-I),
\end{displaymath}

\noindent
where $r$ and $i$ are the instrumental magnitudes of the stars, $R$ and $I$ the reference magnitudes, and $X$
the airmasses (common for both bands) at the time of the observation of the cluster. In our photometric
calibration run we found $r_0=23.01$ mag, $i_0=23.07$ mag, $r_1=0.11$ mag/airmass, $i_1=0.06$ mag/airmass,
$r_2=0.30$ mag and $i_2=0.06$ mag. 
The inverted transformation equations were then used to calibrate the instrumental magnitudes and 
their associated uncertainties of 25 stars in the 0.023 sec exposure, 
325 stars in the 1 sec exposure and 650 stars in the 10 sec observations common in the
I- and R-bands.

%%%%%%%%%%%%%%%%%%%%%%%%%%% FIG. 10
\begin{figure}[t]
        \centering
        \includegraphics[width=8.9cm,angle=0]{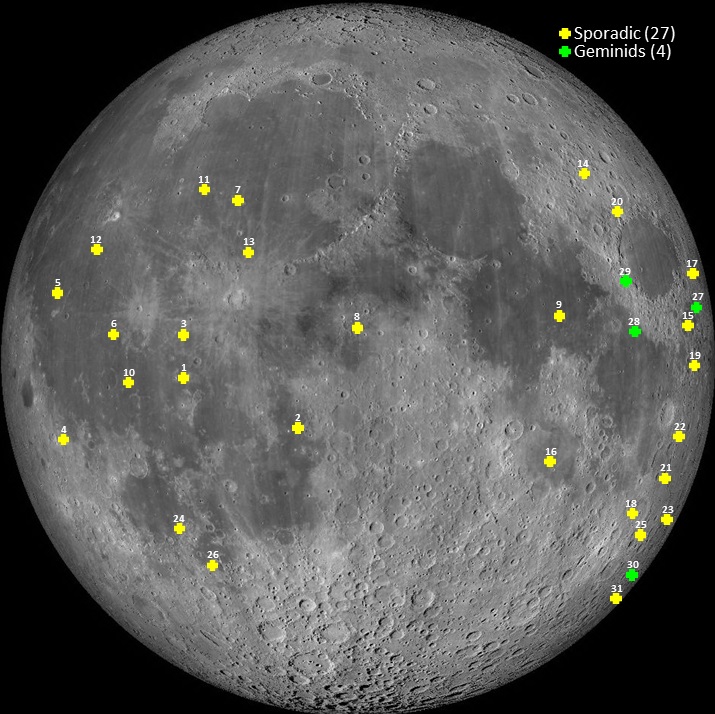}
        \caption{Positions of impacts on the Moon observed during the first year of the
NELIOTA campaign. A total of 31 ``validated'' events have been recorded so far
with 27 being sporadic (yellow points) and 4 possibly originating from the Geminid stream observed
during the period of 12-14 December 2017 (green points).
The index associated with each impact corresponds to the flash ID ($1^{st}$ column in Table~\ref{Tab:Flashes}).
During this period a total of 17 flashes (not shown here) were classified as ``suspected'' since they
were only detected in the I-band.}
        \label{Fig:loc}
\end{figure}

%%%%%%%%%%%%%%%%%%%%%%%%%%% FIG. 11
\begin{figure}[t]
        \centering
        \includegraphics[width=9cm,angle=0]{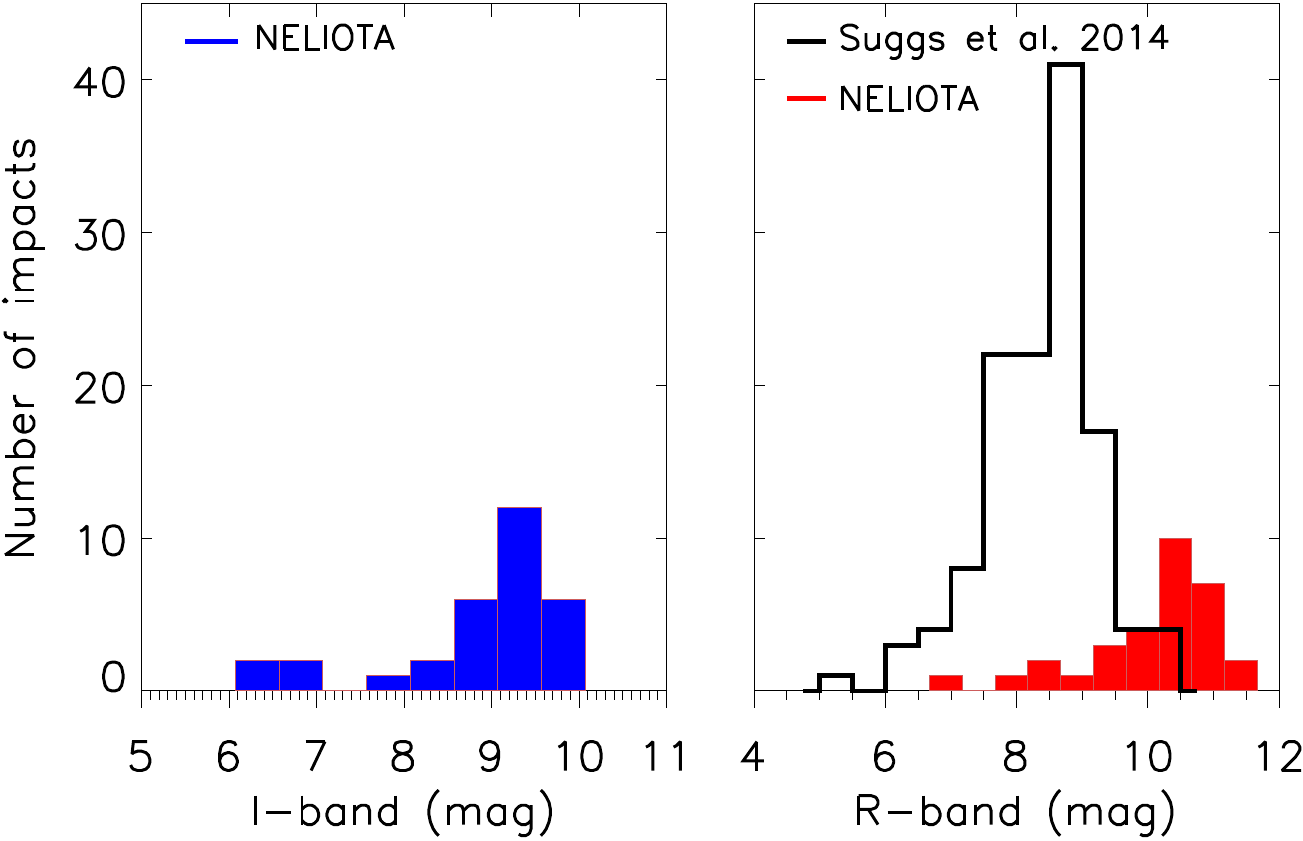}
        \caption{The distribution of the peak magnitudes of the `validated' events in the I-band (left
panel) and in the R-band (right panel). For comparison we have plotted the distribution
of the events observed by \cite{Suggs14} in the R-band (black-line histogram in the
right panel). It is evident that the NELIOTA campaign is detecting flashes
about two magnitudes fainter than what was previously detected by \cite{Suggs14}, a result
of the use of a larger aperture telescope (see text for details). The difference
in the number of the events is mainly due to the different campaign periods
[one year of NELIOTA observations versus five years of observations in \cite{Suggs14}].
}
        \label{Fig:hist}
\end{figure}

The results are shown in Fig.~\ref{Fig:sigmamag}
with the I-band observations in the left panel and the R-band observations 
in the right panel (see the caption of Fig.~\ref{Fig:sigmamag} for the explanation of 
different symbols and lines). For both bands the 0.023 sec, the 1 sec and the 10 sec observations are
shown along with the theoretical predictions imposed by Eq. 1. What is evident is that 
the model predictions follow nicely the measured instrumental magnitudes and associated errors.
Using the model, the magnitudes reached at these exposure times (at a certain noise level) can be derived. 
We find that observations through KPFI of stellar-like objects on a dark night
in the I-band
can reach up to 13.05 mag, 17.07 mag and 18.76 mag in 0.023, 1 and 10 sec, respectively,
at a $0.4\times \sigma_{mag}$ level (corresponding to 0.25 $SNR$ detections). 
At $0.1\times \sigma_{mag}$ level ($SNR=10$) the magnitudes reached are 11.43, 15.45 and 17.21 
in 0.023, 1, and 10 sec, respectively. R-band observations can reach up to 12.81 mag, 16.82 mag 
and 18.74 mag in 0.023, 1 and 10 sec, respectively, at a $0.4\times \sigma_{mag}$ level, while 
the magnitudes that can be reached at a $0.1\times \sigma_{mag}$ level are 11.19, 15.21 and 17.18 in 0.023, 1, 
and 10 sec, respectively.

Additionally to the sources of uncertainties discussed above 
there is another kind of noise 
which becomes significant at short exposures (like in this case) and
can dominate the error in the measurements especially at
large airmasses. This is the scintillation noise ($\sigma_{scint}$) which
depends on many parameters, mainly on the stability of the atmosphere on the night of oservation.
As a first attempt to explore this effect we follow the description described
in \citep{Suggs14} by performing observations of
a photometric standard star (SA~113-475, RA$_{2000}$: $21^h 41^m 51^s$, DEC$_{2000}$: $+00^{\circ} 
39\arcmin 05\arcsec$; \cite{landolt}) on the night of 7 July, 2018. We have observed this
star in both bands at four different airmasses ($X$=1.26, 1.7, 2.2, 2.75) with an exposure time of 23 msec
and for 30 times per airmass. The results of the aperture photometry that we performed,
using the {\it{phot}} task in IRAF$^9$, are shown in the top panel of Fig.~\ref{scintillation}
with the instrumental magnitudes (small circles) in both R- and I-bands (red and blue symbols
respectively) plotted against the airmass. The large circle in each airmass group of points 
is the corresponding mean value of the measured counts in magnitudes.
It is evident that there is a signifficant scatter in brightness,
mainly due to scintillation, around the mean value which gets larger with airmass. This is shown
more clearly in the bottom panel of Fig.~\ref{scintillation} with the scintillation noise ($\sigma_{scint}$),
calculated as the standard deviation of the observations in each airmass group, plotted against
the airmass.
The standard deviation was calculated as the 1-sigma variation of the measured fluxes (in counts).
Linear fits to the data give an estimate of the actual effect of scintillation
with airmass. In particular, we find that in the R-band $\sigma_{scint}$ changes as $0.034+0.026~X$
with airmass and in the I-band this relation becomes $-0.011+0.037~X$. We can see that at large
airmasses this can be the dominant source of uncertainty reaching up to $\sim 0.1$ mag. In the
NELIOTA observations that will follow we will consider a combined photometric and scintillation
error in the form of:
\begin{equation}
\sigma=\sqrt{\sigma_{mag}^2+\sigma_{scint}^2}
\end{equation}
although, in a subsequent paper (Liakos et al. in prep), we plan 
to carry out a more thorough analysis in a subsequent paper
taking into account various seeing conditions and a range in brightness so that more realistic measurements
of the scintillation noise can be obtained.

\subsection{Field distortion}
Despite the simple design of the KPFI, the passage of light through the
optical elements will cause some distortion. The knowledge of the level
of distortion in the two channels of the KPFI is of great importance for
NELIOTA science since it could introduce uncertainties in the
determination of the exact position of the NEO impact. We observed 
the dense stellar field of the open cluster NGC~6811 
(RA$_{2000}$: $19^h 37^m 17^s$, DEC$_{2000}$: $+46^{\circ} 23\arcmin 18\arcsec$),
on the night of October 13$^{th}$, 2017, at 10 sec exposure,
in order to produce the distortion maps in the R- and
I-bands. This was done by calculating the astrometric differences of
the stars in the field using the {\it ccmap} task in IRAF$^7$ 
that provides the plate solution using a list of matched pixel and celestial
coordinates, the latter obtained from \cite{janes}. A total of
30 bright stars located throughout the field were used in both bands 
to compute the plane solution by fitting 3$^{rd}$ order polynomials in both the x- and
y-axis.
The maps are presented in
Fig.~\ref{Fig:distortion} for the I-band (left panel) and for the 
R-band (right panel). The arrows indicate the
expected position of the star (beginning of the arrow) and the observed
position (end of arrow). For displaying purposes the arrows are
artificially enlarged by a factor of 50 with a scale indicator (4 `enlarged' 
pixels) shown in the upper left
part of each plot. It is evident that the distortion is very small in
the center of the field, while it increases towards the edges.
The rms of the fit was found to be 60 mas in the I-band and 55 mas in the R-band.
Nevertheless, the maximum distortion is of the order of the pixel scale and, in any
case, less than the seeing effects, therefore it does not affect the
localization of NELIOTA flashes (accurate to $6.5\arcsec$ on the Moon; see Sect.~\ref{SCIENCE}).

\section{The NELIOTA observations}\label{SCIENCE}

%%%%%%%%%%%%%%%%%%%%%%%%%%% FIG. 12
\begin{figure}[t]
        \centering
        \includegraphics[width=9cm,angle=0]{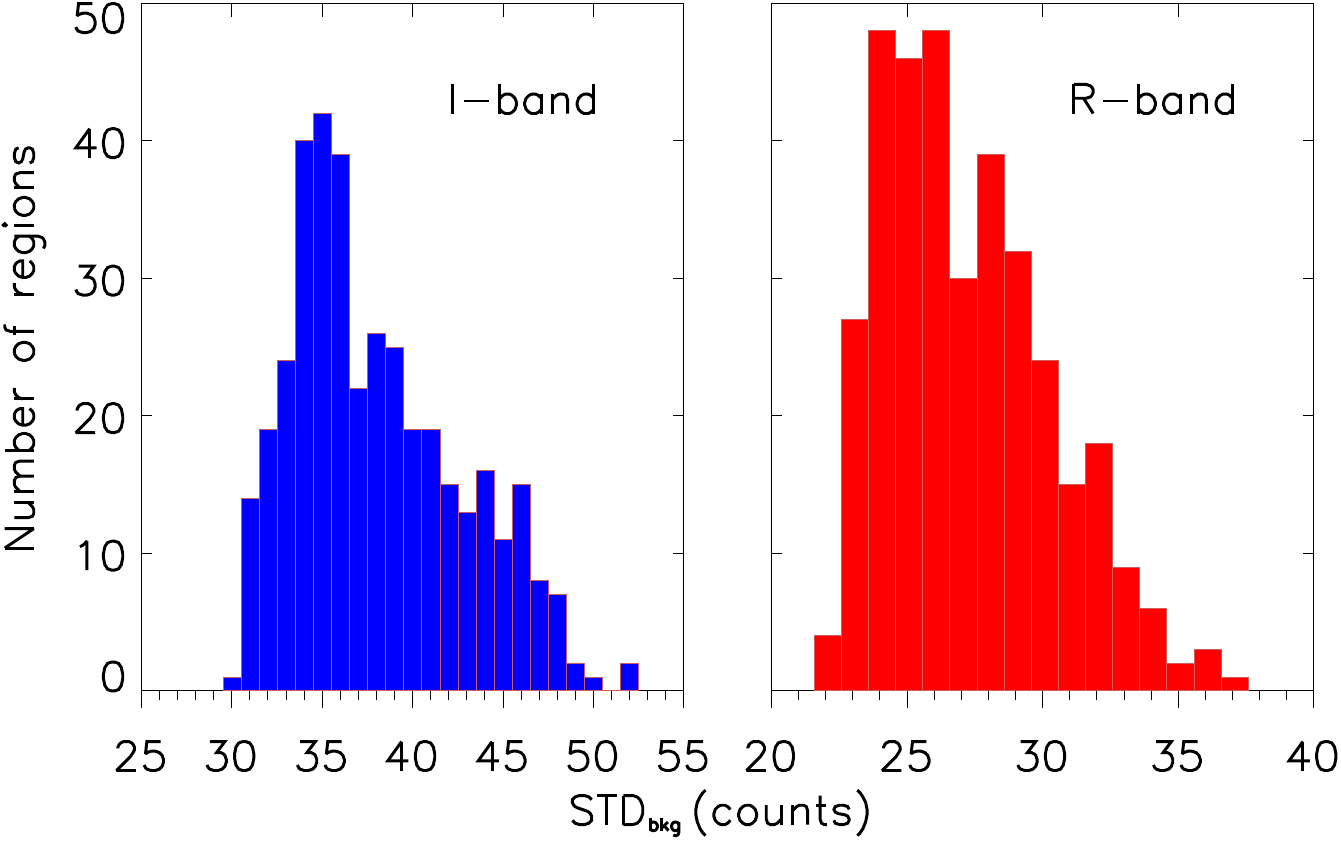}
        \caption{
The background standard deviation ($STD_{bkg}$) distributions in I- and R-band
(left and right panels respectively) for observations made in lunar phase of 0.118 on
March 1$^{st}$, 2017.
Each measurement comes from an aperture of radius of 11 pixels with 352 such regions randomly selected
throughout the Moon's surface observed.
}
        \label{Fig:phase}
\end{figure}
%%%%%%%%%%%%%%%%%%%%%%%%%%% FIG. 13
\begin{figure}[h]
        \centering
        \includegraphics[width=9cm,angle=0]{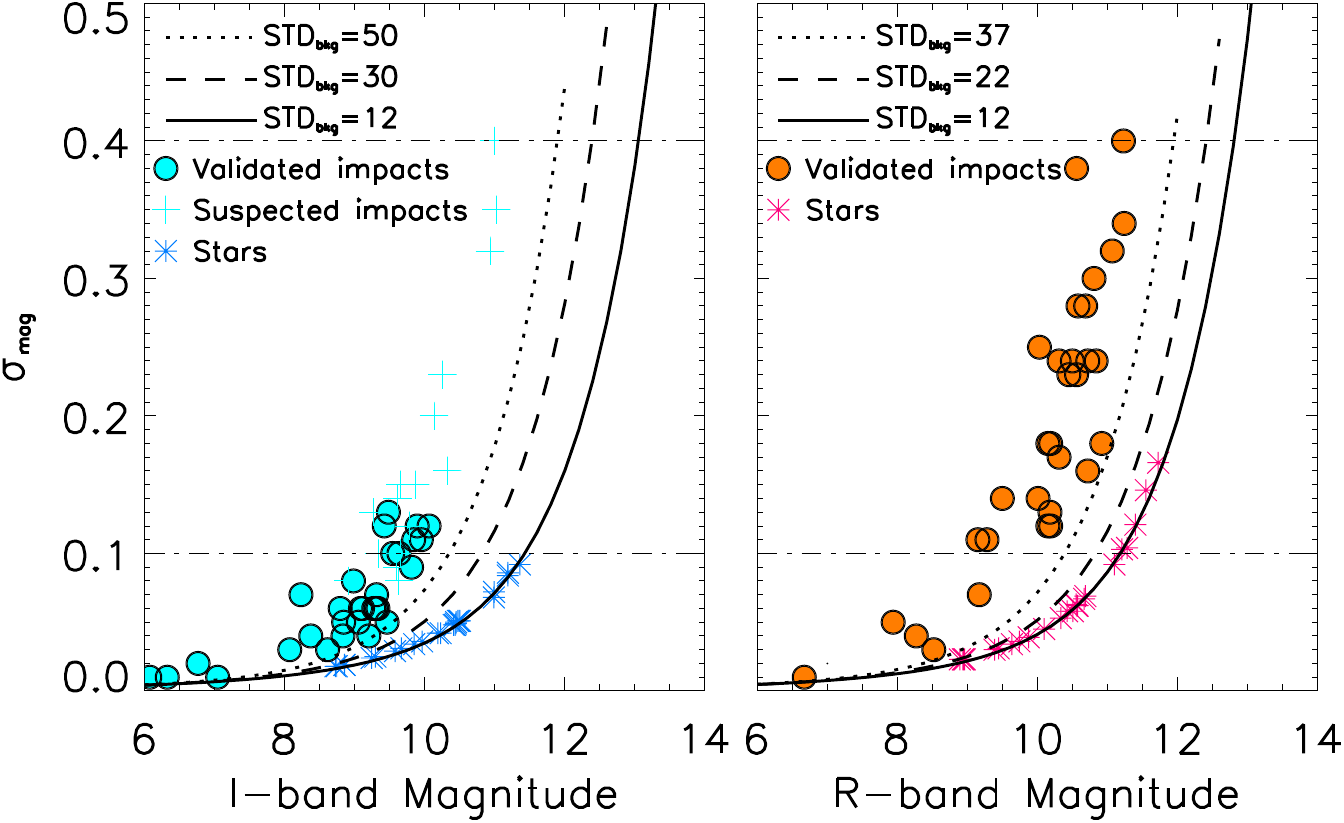}
        \caption{
Measured magnitudes versus the associated magnitude uncertainty ($\sigma_{mag}$) for both stars
and lunar impact flashes in the I- and R-bands (left and right, respectively).
In both panels the stellar measurements are presented as stars, while the 'validated' impact flashes
are shown as circles. For the I-band the
`suspected' impact events (cyan crosses) are also shown.
For the stars, the predicted model (Eq. 1) is also presented (solid line; see Fig.~\ref{Fig:sigmamag}).
The dashed and dotted lines in each panel indicate the model prediction when the minimum and maximum
value, respectively, of the background noise ($STD_{bkg}$) are considered for the case of the observations 
of March 1$^{st}$, 2017 with the lower lunar phase $\sim 0.1$ (Fig.~\ref{Fig:phase}) observed.
}
        \label{Fig:mag_f_s}
\end{figure}

During the first year of the NELIOTA campaign, February 2017\footnote{NELIOTA observations began in 
February during commissioning phase but the official campaign started in March 2017.} - February 2018, 
31 NEO impacts on the Moon have been successfully detected and recorded,
at a frame rate of 30 fps, simultaneously in two photometric bands (R and I; 
Table~\ref{Tab:Flashes}). About half of them 
fade out very fast [within the duration of one single frame (33 msec)], while 
for the rest we were able to follow their brightness variation with time (see Table~\ref{Tab:Flashes}).
An example of such a multi-frame event is shown in Fig.~\ref{Fig:mosaic}.
In this time sequence the light variation of the event observed at 04:35:09.967 UT on December 
14$^{th}$, 2017 (event ID 30) is shown in the four I-band frames in which it was detected. For comparison
the frame just before the event occurred and right after the event finished are also 
presented. The time at the beginning of the exposure in each frame is also given
in each panel (t=0 msec corresponds to the frame where the event was initially detected). 
In one case (event ID 21) the event was captured during its rise with the
I-band brightness measured in the first frame (8.49$\pm$0.04 mag) being slightly lower than
its peak brightness detected in the second frame (8.37$\pm$ 0.04 mag). Similar detections have been 
reported by others \citep{Suggs14} with the video frames showing faint flashes followed by brighter ones.
Photometry of the detected flashes was performed using standard aperture photometry
routines, with a detailed description given elsewhere (Liakos et al., in prep) 
with the uncertainties in brightness calculated as in Eq. 4.
The first science results of the campaign, including determination of the temperatures developed
during the impact events, are presented in \cite{Bonanos17}.

The spatial distribution of the detected impacts on the Moon's surface is 
presented in Fig.~\ref{Fig:loc}. In this figure the location of the 31 validated impacts 
on the lunar surface is indicated with the 27 being sporadic events (yellow symbols) 
and the remaining four (green symbols) related to the Geminid shower observed during 
December 2017 (flashes with IDs 27-30 in Table~\ref{Tab:Flashes}).
Apart from the validated events (observed both in the I- and R-bands) 17 more flashes
have been detected only in the I-band (not shown in Fig.~\ref{Fig:loc}).

%%%%%%%%%%%%%%%%%%%%%%%%%% TABLE 5
\begin{table*}[t]
\tiny
\caption{Basic parameters of the impact flashes observed with the 1.2 m Kryoneri telescope 
as part of the NELIOTA project.}
\begin{center}
\begin{tabular}{lccccccccc}
\hline\hline
%\multicolumn{2}{c}{Virtual server specifications (per server)} \\
Flash ID& Date & UT &Airmass&Lunar phase&Latitude&Longitude & R$\pm \sigma_{\rm R}$& I$\pm \sigma_{\rm I}$ & Duration \\
      &      &  &  &  &  (degrees)& (degrees)&  (mag)      & (mag) & (msec)  \\
\hline
1 & 2017-02-01 & 17:13:57.863&1.66 &0.234& -1.5 &-29.2 &10.15 $\pm$ 0.14 & 9.05 $\pm$ 0.07 & 33 \\
2 & 2017-03-01 & 17:08:46.573&2.18 &0.118& -10.3&-9.7 &6.67 $\pm$ 0.09 & 6.07 $\pm$ 0.07 & 132 \\
3 & 2017-03-01 & 17:13:17.360&2.18 &0.118& 4.5&-29.9 &9.15 $\pm$ 0.14 & 8.23 $\pm$ 0.10 & 33 \\
4 & 2017-03-04 & 20:51:31.853&2.55 &0.412& -12.7&-58.9 &9.50 $\pm$ 0.17 & 8.79 $\pm$ 0.10 & 33 \\
5 & 2017-04-01 & 19:45:51.650&2.44 &0.271&11.6&-58.8 &10.18 $\pm$ 0.16 & 8.61 $\pm$ 0.08 & 33 \\
6 & 2017-05-01 & 20:30:58.137&2.45 &0.350&4.7&-43.2 &10.19 $\pm$ 0.20 & 8.84 $\pm$ 0.09 & 66 \\
7 & 2017-06-27 & 18:58:26.680&2.69 &0.182&26.8&-22.5 &11.07 $\pm$ 0.34 & 9.27 $\pm$ 0.11 & 66 \\
8 & 2017-06-28 & 18:45:25.803&1.93 &0.274&5.6&0.0 &10.56  $\pm$ 0.39 & 9.48$ \pm$ 0.14 & 66 \\
9 & 2017-07-19 & 02:00:36.453&2.31 &0.249&7.8&35.0 &11.23 $\pm$ 0.41 & 9.33 $\pm$ 0.10 & 66 \\
10 & 2017-07-28 & 18:21:44.850&2.04 &0.323&-3.2&-40.0 &11.24 $\pm$ 0.35 &9.29 $\pm$ 0.09 & 66 \\
11 & 2017-07-28 & 18:42:58:027&2.24 &0.323&28.5& -30.6&10.72 $\pm$ 0.26 & 9.63 $\pm$ 0.12 & 33 \\
12 & 2017-07-28 & 18:51:41.683&2.46 &0.323&20.6&-50.7 &10.84 $\pm$ 0.26 & 9.81 $\pm$ 0.12 & 33 \\
13 & 2017-07-28 & 19:17:18.307&2.76 &0.323&18.1&-18.7 & 8.27 $\pm$ 0.11 & 6.32 $\pm$ 0.09 & 165 \\
14 & 2017-08-16 & 01:05:46.763&1.94  &0.393&32.0 &47.5 &10.15$\pm$ 0.20 & 9.54 $\pm$ 0.12 & 66 \\
15 & 2017-08-16 & 02:15:58.813&1.44  &0.393&6.7&68.1 & 10.69$\pm$ 0.29 & 9.11 $\pm$ 0.07 & 66 \\
16 & 2017-08-16 & 02:41:15.113&1.36  &0.393&-15.6&34.6 & 10.81$\pm$ 0.31 & 9.08 $\pm$ 0.07 & 66 \\
17 & 2017-08-18 & 02:02:21.417&2.78  &0.177&-25.9 &57.8 &10.92$\pm$ 0.21 & 9.20 $\pm$ 0.10 & 66 \\
18 & 2017-08-18 & 02:03:08.317&2.78  &0.177&13.5 &76.8 &10.19$\pm$ 0.16 & 8.83 $\pm$ 0.10 & 66 \\
19 & 2017-09-14 & 03:17:49.737&1.17  &0.419&-1.1 &70.0 &9.17$\pm$ 0.10 & 8.07 $\pm$ 0.04 & 99 \\
20 & 2017-09-16 & 02:26:24.933&2.20  &0.200&24.7 &52.5 &8.52$\pm$ 0.10 & 7.04 $\pm$ 0.07 & 231 \\
21 & 2017-10-13 & 01:54:21.710&1.42  &0.449&-17.3&65.2 & 9.28$\pm$ 0.13 & 8.37 $\pm$ 0.06 & 132 \\
22 & 2017-10-13 & 02:33:43.560&1.26  &0.449&-12.5&66.5 & 10.31$\pm$ 0.25 & 9.89 $\pm$ 0.12 & 99 \\
23 & 2017-10-16 & 02:46:45.613&2.98  &0.142&-25.4&72.5 & 10.72$\pm$ 0.20 & 9.46 $\pm$ 0.11 & 99 \\
24 & 2017-10-26 & 17:59:42.880&2.48  &0.390&-27.9&-33.8 & 10.03$\pm$ 0.27 & 9.42 $\pm$ 0.15 & 33 \\
25 & 2017-11-14 & 03:34:15.203&2.08  &0.180&-29.5&64.4 & 10.31$\pm$ 0.19 & 9.31 $\pm$ 0.09 & 66 \\
26 & 2017-11-23 & 16:17:33.000&2.29  &0.227&-35.0&-30.5 & 10.45$\pm$ 0.25 & 10.06 $\pm$ 0.14 & 66 \\
27 & 2017-12-12 & 02:48:08.410&1.92  &0.322&9.0&74.0 & 10.50$\pm$ 0.25 & 8.89 $\pm$ 0.10 & 66 \\
28 & 2017-12-12 & 04:30:00.623&1.41  &0.322&5.4&51.2 & 10.58$\pm$ 0.29 & 9.84 $\pm$ 0.12 & 33 \\
29 & 2017-12-13 & 04:26:57.717&1.72  &0.229&13.0&50.0 & 10.56$\pm$ 0.24 & 9.95 $\pm$ 0.12 & 33 \\
30 & 2017-12-14 & 04:35:09.967&2.14  &0.149&-36.9&73.4 & 7.94$\pm$ 0.10 & 6.76 $\pm$ 0.07 & 132 \\
31 & 2018-01-12 & 03:54:03.027&2.73  &0.204&-40.7&79.2 & 10.01$\pm$ 0.17 & 9.31 $\pm$ 0.11 & 66 \\
\hline
\end{tabular}
\tablefoot{R- and I-band peak magnitudes (columns 8 and 9) and duration of the impact flashes (column 10)
observed with the
1.2 m Kryoneri telescope for the NELIOTA project. The UT date and time at the beginning of
the observation (columns 2 and 3 respectively), the airmass (column 4), the selenographic coordinates 
(Latitude and Longitude; columns 6 and 7) and the lunar phase  (column 5) are also presented.
Each flash is assigned with an ID (column 1). 
The error in the selenographic coordinates
is estimated to be $\sim 0.5$ degrees. Uncertainties in the magnitudes are calculated as in
Eq. 4. Impacts detected in a single frame (in both R- and I-bands) have a maximum duration of 33 msec.}
\end{center}
\label{Tab:Flashes}
\end{table*}

The distribution of the brightness of the flashes detected during the first year 
of the campaign is given in Fig.~\ref{Fig:hist} for the detections in the I-band
(left panel) and the detections in the R-band (right panel). For the multi-frame
events we plot the peak magnitudes which in all cases, except the I-band measurements of event ID 21,
were measured in the frame where the event was first detected.
Something that is directly evident from these histograms is that flashes
are always fainter in the R-band with respect to the I-band. 
An average R-I color index for all the flashes detected by NELIOTA is $1.02\pm0.05$ mag.
For comparison, in Fig.~\ref{Fig:hist} we have also included the R-band magnitude distribution
of the events detected by \cite{Suggs14}.
This shows that the NELIOTA setup was successful in detecting fainter flashes (about 
two magnitudes fainter than was done before). The primary reason for this is the 
larger aperture telescope that is now being used but also the higher efficiency 
detectors that are implemented in the NELIOTA system. Since these fainter 
flashes are more frequent [see, e.g., \cite{Suggs14}] the NELIOTA system is more sensitive in detecting 
those, compared to brighter flashes already detected by other experiments. The difference in
the number statistics between the two samples (\cite{Suggs14} and NELIOTA) comes 
from the fact that the first one accounts for a five year campaign compared to the 
one year operation of NELIOTA.      

Taking into account the total number of impacts (31), the total exposure time during
which these events were collected (54 hours of Moon observations), as well as the average
surface observed per night,
we estimate a detection rate of 1.96$\times10^{-7}$ events km$^{-2}$ h$^{-1}$.
This is about a factor of two higher than what has been reported before
[1.03$\times10^{-7}$ events km$^{-2}$ h$^{-1}$ \citep{Suggs14} and
1.09$\times10^{-7}$ events km$^{-2}$ h$^{-1}$ \citep{Rembold15}].
This is a direct result of the use of a larger aperture telescope (compared to previous
campaigns) allowing for the detection of fainter events (see Fig.~\ref{Fig:hist}), which
are more frequent (see, e.g. \cite{drolshagen} and references therein).

\subsection{Expected limiting magnitudes}
In our attempt to quantify the expected performance of the NELIOTA system, we explored how the 
analysis performed on stellar photometry (Sect.~\ref{noise})
can be applied to predict the limiting magnitudes of the impact flashes that the NELIOTA system 
can reach. One parameter that greately affects the detection of such flashes is
the background noise ($\sigma_{bkg}$; see Eq. 2) which in the case of the Moon, in contrast to the 
uniform sky background observed in stellar objects, varies  significantly depending on the lunar phase.
As the noise of the background ($\sigma_{bkg}$) scales with the square root of the brightness 
of the background, in this case the illumination of the Moon, we expect that $\sigma_{bkg}$
obtains its smallest values on observations obtained when the side of the Moon visible from
the earth is illuminated the least. For NELIOTA scheduling this corresponds to phase $\sim 0.1$.
Even in this case, since the Moon is not uniformly illuminated, $\sigma_{bkg}$                
is expected to vary throughout its surface (gradually becoming brighter from the limb to the center). 
In order to explore the effect of $\sigma_{bkg}$
on the impact flash detections, we calculate the standard deviation ($STD_{bkg}$) in 352, randomly 
selected, regions throughout the Moon's surface, in both bands (R and I). The background was
measured within an aperture of a radius of 11 pixels, which roughly corresponds to an area
similar to the one used to calculate the local background of the stars in the cluster.   
We used the observations performed on March 1$^{st}$, 2017 (exposure time of 23 msec) that corresponds 
to the lowest lunar
phase (0.118) at which obsrevations have been obtained.
The measured values for $STD_{bkg}$ are shown in the two histograms
in Fig.~\ref{Fig:phase} corresponding to the I-band (left panel) and R-band (right panel).
We find that the values of $STD_{bkg}$ range from $\sim 30$ to 50 counts in the I-band,
with the most frequent value at $\sim 35$ counts
and from $\sim 22$ to 37 counts in the R-band, with the most frequent value at $\sim 25$ counts. 

In order to estimate the NELIOTA detection limit we plotted in Fig.~\ref{Fig:mag_f_s} the 
observed magnitudes and their calculated uncertainties for both the impact flashes and 
the stars (see Fig.~\ref{Fig:sigmamag}) at the fixed NELIOTA exposure time of 0.023 sec.
In both bands the observed impact flashes at a given uncertainty level are brighter than their respective 
stellar analogues, a result, mainly due to the significant increase in $\sigma_{bkg}$
compared to the sky background as discussed above. Furthermore, the scatter in brightness of 
the impact flashes at a given uncertainty is, mainly, due to the variation in $\sigma_{bkg}$, which,
as mentioned earlier, greatly depends on the lunar phase. 
In Fig.~\ref{Fig:mag_f_s}, we have also plotted the model predictions (Eq. 1) for the stars
(the solid line; also seen in Fig.~\ref{Fig:sigmamag}), corresponding to a $STD_{bkg}$
value of 12 counts (see Sect.~6.1). Adjusting Eq. 2 for different background noise levels we can then plot the 
model for the range of $STD_{bkg}$ values observed for $\sim 0.1$ lunar phases. We have done
so by plotting the minimum (dashed lines) and maximum (dotted lines) observed values 
of $STD_{bkg}$ (30 and 50 counts in the I-band and 22 and 37 for the R-band).
For the I-band we find that the NELIOTA detections at the $SNR=2.5$
threshold may vary from 11.90 mag to 12.39 mag depending on the measured $STD_{bkg}$ 
(as shown in Fig.~\ref{Fig:phase}). The 10 $SNR$ detections range between 10.35 mag and 10.83 mag.
Similarly, in the R-band, the $SNR=2.5$ detections range between 11.95 mag and 12.41 mag, while
the $SNR=10$ detections range between 10.39 mag and 10.83 mag.

\begin{figure}[t]
        \centering
        \includegraphics[width=9cm,angle=0]{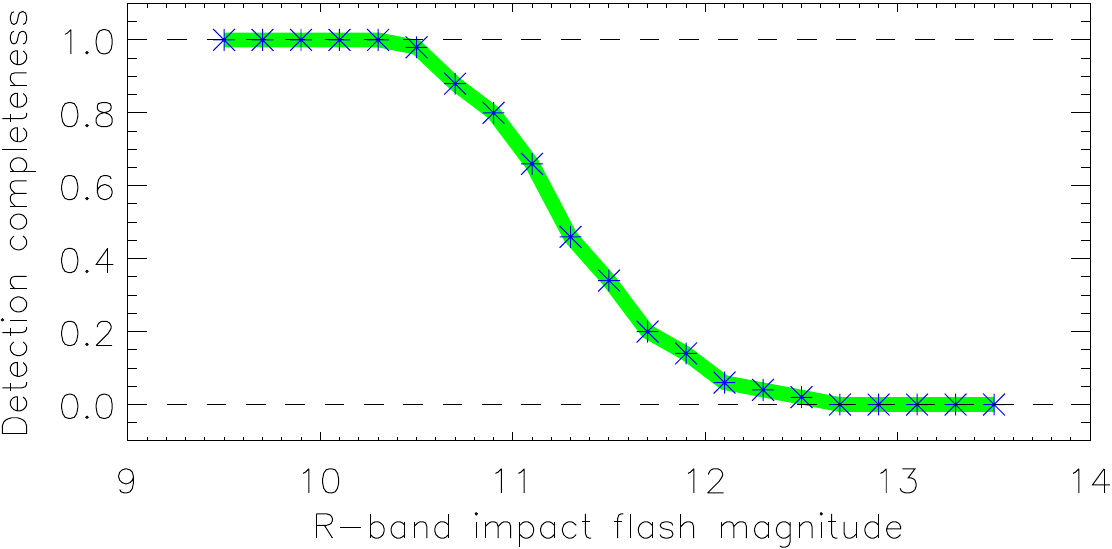}
        \caption{
Detection completeness as a function of the brightness of the impact flashes in the R-band.
For each magnitude bin of 0.2 mag in the R-band 50 mock flashes were placed in random positions 
throughout a real observation of the Moon (in both the R- and I-band images) with a color 
index of R-I=1 mag. The completeness is then computed by forming the ratio of
the ones detected with $SNR$ above 2.5 ($\sigma_{mag}<0.4$), in both R- and I-bands, 
to the total number created (50).
}
        \label{mock}
\end{figure}

We furthermore explored the detection completeness as a funtion of the 
brightness of the impact flash in the R-band. To achieve this we created 
mock impact flashes of certain brightnesses on a real observation of the Moon 
(in both R- and I-band images) and 
subsequently identified the flashes with $SNR$ above 2.5 ($\sigma_{mag}<0.4$). 
A constrain of an average color index of R-I=1, as indicated by our sample 
(see Sect.~\ref{SCIENCE}) was applied.
In more detail, we used the observation of the lowest observed lunar phase of 
March $1^{st}$,2017 (see above) and placed 50 mock flashes in random places throughout the
Moon's surface using the task {\it mkobjects} in IRAF$^9$. Each flash was made to 
resemble point like gaussian distributions of FWHM of 2.4$\arcsec$, the mean seeing value
of all the nights with NELIOTA observations and were grouped in bins of 0.2 mags in
brightness ranging from 9.4 to 13.4 mag in the R-band.
The produced image of the impact flashes on the Moon was then background subtracted with
the procedure described in Sect.~\ref{detection} and used as input to the {\it phot} task in 
IRAF$^9$ in order to count the detections showing a $SNR$ greater than 2.5.  
The completeness was then formed as the ratio of the mock flashes fulfilling the
above criteria to the initial number of flashes created (50).
The results of this procedure are presented in Fig.~\ref{mock} with the detection 
completeness presented as a function of R-band brightness.
The drop in the source completeness begins at 10.4 mag (0.98) and it gets practically
zero (0.02) at 12.4 mag, well in accordance with our previous analysis for the
limiting magnitudes (Fig.~\ref{Fig:mag_f_s}), with a $50\%$ detection completeness 
at 11.4 mag. In a subsequent paper we plan for a more detailed analysis on the
NELIOTA detection completeness taking into account all the parameters that influence such
calculations (e.g. seeing, source morphology, variable background level, and variable 
color index) in order to properly define the intrinsic luminosity function of relatively faint Lunar impact 
flashes.

%%%%%%%%%%%%%%%%%%%%%%%%%%% FIG. 14

\section{Conclusions}\label{SUMMARY}
We present the design and performance of KPFI, a wide-field, high-cadence, twin 
lunar monitoring system at the prime focus of the 1.2~m Kryoneri telescope that 
is used for the NELIOTA project. The project aims at detecting
faint flashes on the Moon's surface produced by impacts of NEOs. The
optical design, the detectors, the control system as well as the
dedicated software for the detection of the NEO impacts are discussed,
while the instrument performance and highlights of the first
scientific results are shown. The novelty of the NELIOTA system 
is the use of a large aperture telescope (larger than ever used before
for this purpose) and the high-cadence achieved by two detectors
observing simultaneously at a rate of 30 frames-per-second in two optical bands
(R and I).  
The noise model predicts that the NELIOTA system can detect NEO impact flashes
at a $SNR=2.5$ level of 12.39 mag in the I-band and 12.41 mag in the R-band for observations 
made at low lunar phase ($\sim 0.1$).
During the first year of operation, 31 such NEO impacts on the Moon's surface were observed 
with the faintest flash being 11.24 mag in the R-band
(about two magnitudes fainter than ever observed before) translated into a detection 
rate of 1.96$\times 10^{-7}$ events km$^{-2}$ h$^{-1}$. The ability to monitor the impact flashes
in two bands at the same time provides unique and significant constraints on the temperatures 
produced during the impact.

The wide-field, high-cadence and simultaneous multicolor abilities of KPFI 
make it a unique instrument that can be used by the community for
a variety of astronomy projects
such as occultations (e.g. Sicardy et al., in prep), exoplanet transit 
light curves, monitoring of early supernova light curves [e.g. \cite{Bonanos16}], as well as transient 
follow up [e.g. \cite{Wyrz}] and other NEO science.

The reader is encouraged to visit the dedicated NELIOTA website$^1$ for further information.

\begin{acknowledgements} %%%%%%%%%%%%%%%%%%%%%%%%%%%%%%%%%%%%%%%%%%%%%%%%  
The authors wish to thank the anonymous referee and the Editor (Thierry Forveille)
for constructive comments that helped improving the paper.
We are also greatful to A. Georgakakis for stimulating discussions and to E. Palaiologou 
for useful guidance on the field distortion calculations.
The authors greatfully acknowledge financial support by the European
Space Agency under the NELIOTA program, contract No. 4000112943. This
work has made use of data from the European Space Agency NELIOTA
project, obtained with the 1.2~m Kryoneri telescope, which is operated
by IAASARS, National Observatory of Athens, Greece.
\end{acknowledgements}

\bibliography{NELIOTA} % your references Yourfile.bib

\begin{thebibliography}{27}
\expandafter\ifx\csname natexlab\endcsname\relax\def\natexlab#1{#1}\fi

\bibitem[{{Ait Moulay Larbi} {et~al.}(2015){Ait Moulay Larbi}, {Daassou},
  {Baratoux}, {Bouley}, {Benkhaldoun}, {Lazrek}, {Garcia}, \&
  {Colas}}]{AitMoulay15}
{Ait Moulay Larbi}, M., {Daassou}, A., {Baratoux}, D., {et~al.} 2015, Earth
  Moon and Planets, 115, 1

\bibitem[{{Artem'eva} {et~al.}(2001){Artem'eva}, {Kosarev}, {Nemtchinov},
  {Trubetskaya}, \& {Shuvalov}}]{Artemeva01}
{Artem'eva}, N.~A., {Kosarev}, I.~B., {Nemtchinov}, I.~V., {Trubetskaya},
  I.~A., \& {Shuvalov}, V.~V. 2001, Solar System Research, 35, 177

\bibitem[{{Bellot Rubio} {et~al.}(2000){Bellot Rubio}, {Ortiz}, \&
  {Sada}}]{BellotRubio00}
{Bellot Rubio}, L.~R., {Ortiz}, J.~L., \& {Sada}, P.~V. 2000, \apjl, 542, L65

\bibitem[{{Bonanos} {et~al.}(2018){Bonanos}, {Avdellidou}, {Liakos},
  {Xilouris}, {Dapergolas}, {Koschny}, {Bellas-Velidis}, {Boumis},
  {Charmandaris}, {Fytsilis}, \& {Maroussis}}]{Bonanos17}
{Bonanos}, A.~Z., {Avdellidou}, C., {Liakos}, A., {et~al.} 2018, \aap, 612, A76

\bibitem[{{Bonanos} \& {Boumis}(2016)}]{Bonanos16}
{Bonanos}, A.~Z. \& {Boumis}, P. 2016, \aap, 585, A19

\bibitem[{{Bouley} {et~al.}(2012){Bouley}, {Baratoux}, {Vaubaillon}, {Mocquet},
  {Le Feuvre}, {Colas}, {Benkhaldoun}, {Daassou}, {Sabil}, \&
  {Lognonn{\'e}}}]{Bouley12}
{Bouley}, S., {Baratoux}, D., {Vaubaillon}, J., {et~al.} 2012, \icarus, 218,
  115

\bibitem[{{Brown} {et~al.}(2013){Brown}, {Assink}, {Astiz}, {Blaauw},
  {Boslough}, {Borovi{\v c}ka}, {Brachet}, {Brown}, {Campbell-Brown},
  {Ceranna}, {Cooke}, {de Groot-Hedlin}, {Drob}, {Edwards}, {Evers}, {Garces},
  {Gill}, {Hedlin}, {Kingery}, {Laske}, {Le Pichon}, {Mialle}, {Moser},
  {Saffer}, {Silber}, {Smets}, {Spalding}, {Spurn{\'y}}, {Tagliaferri}, {Uren},
  {Weryk}, {Whitaker}, \& {Krzeminski}}]{Brown13}
{Brown}, P.~G., {Assink}, J.~D., {Astiz}, L., {et~al.} 2013, \nat, 503, 238

\bibitem[{{Drolshagen} {et~al.}(2017){Drolshagen}, {Koschny}, {Drolshagen},
  {Kretschmer}, \& {Poppe}}]{drolshagen}
{Drolshagen}, G., {Koschny}, D., {Drolshagen}, S., {Kretschmer}, J., \&
  {Poppe}, B. 2017, \planss, 143, 21

\bibitem[{{Everett} \& {Howell}(2001)}]{everett}
{Everett}, M.~E. \& {Howell}, S.~B. 2001, \pasp, 113, 1428

\bibitem[{{Goode} {et~al.}(2001){Goode}, {Qiu}, {Yurchyshyn},
  {et~al.}}]{Goode01}
{Goode}, P.~R., {Qiu}, J., {Yurchyshyn}, V., {et~al.} 2001, \grl, 28, 1671

\bibitem[{{Harris} \& {D'Abramo}(2015)}]{Harris15}
{Harris}, A.~W. \& {D'Abramo}, G. 2015, \icarus, 257, 302

\bibitem[{{Ivanov} {et~al.}(2002){Ivanov}, {Neukum}, {Bottke}, \&
  {Hartmann}}]{Ivanov02}
{Ivanov}, B.~A., {Neukum}, G., {Bottke}, Jr., W.~F., \& {Hartmann}, W.~K. 2002,
  {The Comparison of Size-Frequency Distributions of Impact Craters and
  Asteroids and the Planetary Cratering Rate}, ed. W.~F. {Bottke}, Jr.,
  A.~{Cellino}, P.~{Paolicchi}, \& R.~P. {Binzel}, 89--101

\bibitem[{{Janes} {et~al.}(2013){Janes}, {Barnes}, {Meibom}, \& {Hoq}}]{janes}
{Janes}, K., {Barnes}, S.~A., {Meibom}, S., \& {Hoq}, S. 2013, \aj, 145, 7

\bibitem[{{Jeffries} {et~al.}(2013){Jeffries}, {Naylor}, {Mayne}, {Bell}, \&
  {Littlefair}}]{jeff}
{Jeffries}, R.~D., {Naylor}, T., {Mayne}, N.~J., {Bell}, C.~P.~M., \&
  {Littlefair}, S.~P. 2013, \mnras, 434, 2438

\bibitem[{{Landolt}(1992)}]{landolt}
{Landolt}, A.~U. 1992, \aj, 104, 340

\bibitem[{{Madiedo} {et~al.}(2014){Madiedo}, {Ortiz}, {Morales}, \&
  {Cabrera-Ca{\~n}o}}]{Madiedo14}
{Madiedo}, J.~M., {Ortiz}, J.~L., {Morales}, N., \& {Cabrera-Ca{\~n}o}, J.
  2014, \mnras, 439, 2364

\bibitem[{{Moser} {et~al.}(2011){Moser}, {Suggs}, {Swift}, {Suggs}, {Cooke},
  {Diekmann}, \& {Koehler}}]{Moser11}
{Moser}, D.~E., {Suggs}, R.~M., {Swift}, W.~R., {et~al.} 2011, in Meteoroids:
  The Smallest Solar System Bodies, ed. W.~J. {Cooke}, D.~E. {Moser}, B.~F.
  {Hardin}, \& D.~{Janches}, 142

\bibitem[{{Nemtchinov} {et~al.}(1998){Nemtchinov}, {Shuvalov}, {Artem'eva},
  {Ivanov}, {Kosarev}, \& {Trubetskaya}}]{Nemtchinov98}
{Nemtchinov}, I.~V., {Shuvalov}, V.~V., {Artem'eva}, N.~A., {et~al.} 1998,
  Solar System Research, 32, 99

\bibitem[{{Ortiz} {et~al.}(2015){Ortiz}, {Madiedo}, {Morales}, {Santos-Sanz},
  \& {Aceituno}}]{Ortiz15}
{Ortiz}, J.~L., {Madiedo}, J.~M., {Morales}, N., {Santos-Sanz}, P., \&
  {Aceituno}, F.~J. 2015, \mnras, 454, 344

\bibitem[{{Ortiz} {et~al.}(2000){Ortiz}, {Sada}, {Bellot Rubio}, {Aceituno},
  {Aceituno}, {Guti{\'e}rrez}, \& {Thiele}}]{Ortiz00}
{Ortiz}, J.~L., {Sada}, P.~V., {Bellot Rubio}, L.~R., {et~al.} 2000, \nat, 405,
  921

\bibitem[{{Qiu} {et~al.}(2013){Qiu}, {Mao}, {Lu}, {Xiang}, \&
  {Jiang}}]{Qiu2013}
{Qiu}, P., {Mao}, Y.-N., {Lu}, X.-M., {Xiang}, E., \& {Jiang}, X.-J. 2013,
  Research in Astronomy and Astrophysics, 13, 615

\bibitem[{{Rembold} \& {Ryan}(2015)}]{Rembold15}
{Rembold}, J.~J. \& {Ryan}, E.~V. 2015, \planss, 117, 119

\bibitem[{{Robinson} {et~al.}(2015){Robinson}, {Boyd}, {Denevi},
  {et~al.}}]{Robinson15}
{Robinson}, M.~S., {Boyd}, A.~K., {Denevi}, B.~W., {et~al.} 2015, \icarus, 252,
  229

\bibitem[{{Speyerer} {et~al.}(2016){Speyerer}, {Povilaitis}, {Robinson},
  {Thomas}, \& {Wagner}}]{Speyerer16}
{Speyerer}, E.~J., {Povilaitis}, R.~Z., {Robinson}, M.~S., {Thomas}, P.~C., \&
  {Wagner}, R.~V. 2016, \nat, 538, 215

\bibitem[{{Suggs} {et~al.}(2014){Suggs}, {Moser}, {Cooke}, \&
  {Suggs}}]{Suggs14}
{Suggs}, R.~M., {Moser}, D.~E., {Cooke}, W.~J., \& {Suggs}, R.~J. 2014,
  \icarus, 238, 23

\bibitem[{{Werner} {et~al.}(2002){Werner}, {Harris}, {Neukum}, \&
  {Ivanov}}]{Werner02}
{Werner}, S.~C., {Harris}, A.~W., {Neukum}, G., \& {Ivanov}, B.~A. 2002,
  \icarus, 156, 287

\bibitem[{{Wyrzykowski} {et~al.}(2017){Wyrzykowski}, {Mroz}, {Rybicki},
  {Altavilla}, {Bakis}, {Bendjoya}, {Birenbaum}, {Blagorodnova},
  {Blanco-Cuaresma}, {Bonanos}, {Bozza}, {Britavskiy}, {Burgaz}, {Butterley},
  {Capuozzo}, {Carrasco}, {Chruslinska}, {Damljanovic}, {Dapergolas},
  {Dennefeld}, {Dhillon}, {Dominik}, {Esenoglu}, {Fossey}, {Gomboc},
  {Hallokoun}, {Hamanowicz}, {Hardy}, {Hudec}, {Khamitov}, {Klencki},
  {Kolaczkowski}, {Kolb}, {Leonini}, {Leto}, {Lewis}, {Liakos}, {Littlefair},
  {Maoz}, {Maund}, {Mikolajczyk}, {Palaversa}, {Pawlak}, {Penny}, {Piascik},
  {Reig}, {Rhodes}, {Russell}, {Sanchez}, {Shappee}, {Shvartzvald}, {Sitek},
  {Sniegowska}, {Sokolovsky}, {Steele}, {Street}, {Tomasella}, {Trascinelli},
  {Wiersema}, {Wilson}, {Zharkov}, {Zola}, \& {Zubareva}}]{Wyrz}
{Wyrzykowski}, L., {Mroz}, P., {Rybicki}, K., {et~al.} 2017, The Astronomer's
  Telegram, 10341

\end{thebibliography}

\end{document}